\documentclass[12pt]{article}

\usepackage{scicite}

\usepackage{graphicx}
\usepackage{times}

\newenvironment{sciabstract}{%
\begin{quote} \bf}
{\end{quote}}

\newcounter{lastnote}

\usepackage{scicite,amsmath}

\usepackage{color}

\usepackage{graphicx}
\newcommand{\upperRomannumeral}[1]{\uppercase\expandafter{\romannumeral#1}}

\usepackage{times}

\title{Quantized Faraday and Kerr rotation and axion electrodynamics of a 3D topological insulator}

\author
{Liang Wu,$^{1\ast, \dag}$ M. Salehi,$^{2}$ N. Koirala,$^{3}$  J. Moon,$^{3}$   \\ S. Oh,$^3$ N. P. Armitage,$^{1\ast}$\\
\\
\normalsize{$^{1}$The Institute for Quantum Matter, Department of Physics and Astronomy,} \\
\normalsize{ The Johns Hopkins University, Baltimore, MD 21218 USA.}\\
\normalsize{$^{2}$Department of Material Science and Engineering,} \\
 \normalsize{Rutgers the State University of New Jersey. Piscataway, NJ 08854}\\
\normalsize{$^{3}$Department of Physics and Astronomy,} \\
\normalsize{ Rutgers the State University of New Jersey. Piscataway, NJ 08854}\\
\\
\normalsize{$^\ast$To whom correspondence should be addressed; E-mail:  liangwu@berkeley.edu; npa@jhu.edu}\\
\normalsize{$^\dag$Present address: Department of Physics, University of California, Berkeley. CA 94720}
}

\date{\today}

\begin{document} 


\baselineskip24pt


\maketitle


\begin{sciabstract}
 Topological insulators have been proposed to be best characterized as bulk magnetoelectric materials that show response functions quantized in terms of fundamental physical constants. Here we lower the chemical potential of three-dimensional (3D) Bi$_2$Se$_3$ films to $\sim$ 30 meV above the Dirac point, and probe their low-energy electrodynamic response in the presence of magnetic fields with high-precision time-domain terahertz polarimetry. For fields higher than 5 T, we observed quantized Faraday and Kerr rotations, whereas the DC transport is still semi-classical. A non-trivial Berry phase offset to these values gives evidence for axion electrodynamics and the topological magnetoelectric effect.  The time structure used in these measurements allows a direct measure of the fine structure constant based on a topological invariant of a solid-state system.

\end{sciabstract}

Topological phenomena in condensed matter physics provide some of the most precise measurements of fundamental physical constants.  The measurement of the quantum conductance $G_{0}=e^{2}/h$ from the quantum Hall effect \cite{KlitzingPRL1980} and the flux quantum $\Phi_{0}=h/2e$ from the Josephson effect \cite{josephson1962possible, anderson1963probable} provide the most precise value for Planck's constant $h$. In the last number of years a new class of materials in the form of topological insulators has been discovered \cite{moore2010birth, HasanKaneRMP10, Qi-Zhang-11}.   These are materials in which topological properties of the bulk wavefuctions give rise to a topologically protected surface metal with a novel massless Dirac spectrum.   It has been proposed that topological insulators are best characterized not as \textit{surface} conductors, but as \textit{bulk} magnetoelectrics \cite{Qi08b, Essin09a} with a quantized magnetoelectric response coefficient whose size is set by the fine structure constant $\alpha=e^2/2 \epsilon_0 hc $.  Such a topological measurement could provide precise values for three fundamental physical constants: the electric charge $e$, Planck's constant $h$, and the vacuum impedance  $Z_0 = \sqrt{\mu_0 / \epsilon_0}$ in a solid-state context.

Magnetoelectrics are materials in which a polarization can be created by an applied magnetic field or a magnetization can be created by an applied electric field and have been topics of interest for decades \cite{fiebig2005revival}.  Representative examples of magnetoelectric (ME) materials are Cr$_2$O$_3$ \cite{dzyaloshinskii1959magneto} with an $\textbf{E} \cdot \textbf{B}$ ME coupling of the form  and multiferroic BiFeO$_3$ \cite{wojdel2009magnetoelectric} which can be written (in part) as an $\textbf{E} \times \textbf{B}$ ME coupling.     Topological insulators can be characterized as special $\textbf{E} \cdot \textbf{B}$ magnetoelectrics \cite{Qi08b, Essin09a}, which in the topological field theory can be shown to be a consequence of an additional term $\mathcal{L}_{\theta} =  - 2 \alpha \sqrt{ \frac{\epsilon_0}{\mu_0} }   \frac{ \theta}{ 2 \pi}   \textbf{E} \cdot  \textbf{B}$ added to the usual Maxwell Lagrangian \cite{Qi08b}.  Here $\alpha$ is the fine structure constant, and $\epsilon_0$ and $\mu_0$ are the permittivity and permeability of free space.   

Although $\mathcal{L}_{\theta} $ is generic expression which  can be applied for instance even to Cr$_2$O$_3$ (with $\theta \approx \frac{\pi}{36}$ at low temperature \cite{Hehl2009a}) its form merits additional discussion when applied to TIs.   Although it is usually said that one must break both TRS and inversion to define a magnetoelectric coefficient, this is not precisely true.   The Lagrangian defines the action  $\mathcal{S}  =  \int dt dx^3 \mathcal{L} $ and since all physical observables depend on exp$(i \mathcal{S} / \hbar)$ they are invariant to global changes to $\theta$ of $2\pi$.   Therefore due to the transformation properties of $ \textbf{E}$ and  $\textbf{B}$, if either TRS or inversion are present,  $\theta$ is constrained to be not just zero (as it is in a non-mangetoelectric conventional material), but can take on integer multiples of $\pi$.  Three-dimensional insulators in which either TRS or inversion symmetry is preserved can be divided into two classes of materials, in which $\theta$ is either $2 \pi (N+\frac{1}{2})$ (topological)  or $2 \pi (N)$  (conventional) \cite{Qi08b}.  Here $N$ is an integer that indicates the highest fully filled Landau level (LL) of the surface if TRS is broken.  In either case, $\theta$ can be formulated as a bulk quantity modulo a quantum (here $2 \pi$) in much the same way as the electric polarization  \textbf{P} in a ferroelectric can only be defined as a bulk quantity moduluo a dipole quantum that depends on the surface charge \cite{KingSmith93a}.  It is important to note that in order to support a macroscopic magnetic/electric moment of the sample from an applied electric/magnetic field, macroscopic TRS and inversion must both be broken (as they are in conventional magnetoelectrics) but a finite magnetoelectric $\textbf{E} \cdot \textbf{B}$ term is more general than the capacity to support a moment.  Because inversion symmetric Bi$_2$Se$_3$ in magnetic field breaks only TRS, such a sample cannot exhibit a net macroscopic moment from magnetoelectricity unless inversion is broken macroscopically through some other means, for instance by putting the sample in an $\textbf{E} $ field that preferentially charge biases the surfaces or depositing a magnetic layer where everywhere the magnetization points outwards.   In the case relevant for our experiment, inversion symmetry constrains the crystal's bulk $\theta$ term to be $2 \pi (N+\frac{1}{2})$.   A net macroscopic moment cannot be generated, but the sample is still magnetoelectric in the sense that $\mathcal{L}_{\theta} $ still applies.

An analogy can be made between the physics described here to that of the hypothetical axion particle that was proposed to explain charge conjugation parity symmetry violation (CP violation) in the strong interaction by Wilczek \cite{Wilczek87a} and hence a topological magnetoelectric effect (TME) of this kind has been called `axion electrodynamics'. The axion has not been observed in particle physics experiments, but one may study the analogous effect in the context of topological insulators.

In the limit where a TRS breaking field is small, but finite, and the surface chemical potential is tuned near the Dirac point a modified Maxwell's equations can be derived (see supplementary information (SI) section  \textbf{1}  ) from the full Langragian.   The modified Gauss's and Amp\`ere's laws read

\begin{equation}
\nabla \cdot   \mathbf{E} = \frac{\rho}{\epsilon_0} -  2 c \alpha  \nabla (\frac{\theta}{2 \pi}) \cdot   \mathbf{B}.
\label{Gauss}
\end{equation}

\begin{equation}
\nabla \times    \mathbf{B} = \mu_0   \mathbf{J}  + \frac{1}{c^2}  \frac{\partial   \mathbf{E}}{\partial    t} +  \frac{2 \alpha}{c} [  \mathbf{B}   \frac{\partial }{\partial    t}   (  \frac{\theta}{2 \pi} )  + \nabla (\frac{\theta}{2 \pi}) \times    \mathbf{E}  ].
\label{Ampere}
\end{equation}

The consequences of axion electrodynamics  are the additional source and current terms related to the derivatives of $\theta$ in these modified Maxwell's equations.   The additional source term can lead to an effect whereby an electric charge placed near the surface of a TI creates an image magnetic monopole \cite{qi2009inducing}.   The additional current term gives a novel half-integer quantum Hall effect (QHE) on the TI surface \cite{Qi08b}.  Although there has been indirect evidence for half-integer QHE effects in gated TI BiSbTeSe$_2$ exfoliated flakes \cite{XuNatPhys14},  gated (Bi$_{1-x}$Sb$_x$)$_2$Te$_3$ thin films \cite{yoshimiNC15} and surface charge-transfer doped pure Bi$_2$Se$_3$ films \cite{KoiralaNL2015} at very high magnetic fields, it is generally not straightforward to observe the QHE in a conventional transport style experiment as one does with a 2D electron gas (2DEG) with leads connected to sample edges as TIs have a closed surface with no boundaries{\cite{vafek2011quantum}.  It is desirable then to use non-contact probes such as Faraday and Kerr rotations \cite{Qi08b, Maciejko10a, Tse10a}, which have been predicted to be quantized in units of the fine structure constant.  One can proceed from the modified Amp\`ere's law Eq. \ref{Ampere} in conjunction with the usual Faraday's law to derive the reflection and transmission coefficients for a traveling wave incident on a TI surface (See SI section \textbf{2} ). In an applied magnetic field, one finds that for a TI thin film on a simple dielectric substrate, the Faraday rotation in the quantum regime is

\begin{equation}
\tan(\phi_F)=\frac{2\alpha}{1+n}(N_t+\frac12+N_b+\frac12),
\label{eq3}
\end{equation}

\noindent where $n \sim$ 3.1 is the THz range index of refraction of the sapphire and $N_t$, $N_b$ are the highest fully filled LL of the top and bottom surfaces, which depend on the chemical potential and size of the TRS breaking field.  The effect of considering a TRS breaking field that is oppositely directed with respect to the surface normal for our film's two surfaces gives Eq. \ref{eq3}. 

There have been a number of inter-related challenges in realizing the TME on both the materials and instrumentation side.   First, one must have a negligible level of bulk carriers and a low chemical potential at the surface, but most known topological insulators suffer from inadvertent bulk doping.   A metallic gate used to gate away charge carriers in transport experiments cannot be used easily in an optical experiment because it would have its own Faraday effect.  Second, as the topological field theory is derived for the translationally invariant case, one may expect that it will only apply when the TRS breaking perturbation is strong enough to overcome disorder and establish a surface QHE.  Third, to reveal the TME it is required that the probe frequencies and temperatures be well below the Landau level spacing of the surface states, which are given by $E = \textbf{v}_F  \sqrt{2 N e \mathbf{B} \hbar }$ (where   $\textbf{v}_F$ is the Fermi velocity).  In practice this put the relevant frequency in the sub-THz part of the electromagnetic spectrum, which has been a traditionally difficult frequency range to characterize materials in.   Fourth, THz range experiments with their long wavelengths require large uniform samples of at least a few mm in spatial extent.  Fifth, as the size of the effect is set by the fine structure constant, the expected rotations are expected to be very small and well beyond the capacity of conventional THz range polarimetry.

Despite these challenges, due to a number of innovations, we have succeeded in measuring the quantized Faraday and Kerr rotation using time-domain THz spectroscopy (TDTS)  (1 THz $\sim$ 4.14 meV), using recently developed low-density and high-mobility Bi$_2$Se$_3$ molecular beam epitaxy films, in conjunction with a novel high-precision polarimetry technique \cite{MorrisOE12}.   TDTS is a powerful tool to study the long-wavelength (low-energy) electrodynamics of topological insulators.   Samples are thin films of Bi$_2$Se$_3$ grown by molecular beam epitaxy with a recently developed recipe that results in true bulk insulating TIs with low surface chemical potential.   These films were further treated $in$ $situ$ by a thin charge transfer layer of deposited MoO$_3$ that further decreases the carrier density and put the chemical potential in the bulk gap.  MoO$_3$ is semiconductor with a gap $\sim$ 3 eV \cite{hussain2001optical} and does not contribute to Faraday rotation. Details of the film growth can be found in Ref. \cite{KoiralaNL2015}.

In Fig. \ref{Fig1}(a), we show the real part of the conductance of new generation Bi$_2$Se$_3$ films with and without the MoO$_3$ layer in zero magnetic field as compared to the conductance of an older (but still excellent \cite{ValdesAguilarPRL12, WuNatPhys13}) generation of Bi$_2$Se$_3$ films. As shown in previous work \cite{ValdesAguilarPRL12, WuNatPhys13}, the complex conductance can be characterized by a low-frequency Drude term peaked at zero frequency, a phonon term peaked at $\sim$ 1.9 THz and a lattice polarizability contribution $\epsilon_{\infty}$ (See SI section \textbf{4}  for details). The total area below the Drude conductance is the Drude spectral weight and is proportional to the carrier density ratioed to an effective transport mass.   In previous work \cite{ValdesAguilarPRL12, WuNatPhys13}, we have shown that this spectral weight is independent of film thickness establishing a principle surface transport channel.  In these new Bi$_2$Se$_3$ films with MoO$_3$  there is a weak dependence of this spectral weight on thickness below 10 QL as shown in SI section \textbf{4}.   We believe this is because for this generation of samples, the substantial dopants come from bulk defects (as opposed to surface defects in earlier generation samples) that move to the surface and work as surface carriers.   Additionally it is likely that MoO$_3$ is more effective in reducing carrier density in thinner samples. 

As a part of the magneto-THz experiments described below, we measure the low field cyclotron resonance frequency and find the effective mass using the relation $\frac{\omega_c}{2\pi} = \frac{e B}{m^*}$ (See SI section \textbf{4} for raw data). As show in Fig. \ref{Fig1}(b), a decreased effective mass from 0.15 m$_e$ to 0.07 m$_e$ is observed through charge transfer to MoO$_3$, which is consistent with the known band structure for Dirac surface states, i.e. the dependence of the cyclotron mass is proportional to the square root of the carrier density (See SI section \textbf{4} for more details).  Both bulk states or a trivial 2DEG from an accumulation layer will exhibit a carrier-density independent mass.  Moreover 0.07 m$_e$ is lower than any reported bulk or 2DEG effective mass. Converting either the spectral weight or effective mass to chemical potential using the known band structure of Bi$_2$Se$_3$ these new films with MoO$_3$ have $E_F$ of 30 - 60 meV above the Dirac point and in the bulk gap when an equal contribution from the two surface states is assumed.   Using this mass and the spectral weight by fitting the Faraday rotation, also allows us to directly measure the charge density and quantify (under applied magnetic field) the filling factor $\nu$ (See SI section \textbf{4}).

Having identified the topological surface states and verified the low Fermi energy of these films, we explore their low-frequency Faraday rotation. The experimental configuration is shown in Fig. \ref{Fig2}(a). We measure complex Faraday rotation through TDTS with the polarization modulation technique \cite{MorrisOE12, WuPRL2015}. The Faraday rotation is a complex quantity in which the real part is the rotation of the major axis of the ellipse and the imaginary part is related to the ellipticity as shown in Fig. \ref{Fig2}(b).  The full-field data of a 10 QL sample is shown in Fig. \ref{Fig2}(c) (d). At low fields ($\textless$ 4 T), the Faraday rotation shows semi-classical cyclotron resonance, as demonstrated by the shifting of the inflection point (close to the zero value) in the real part and the shifting of the minimum in the imaginary part with fields as discussed in detail in Ref. \cite{WuPRL2015}. The cyclotron frequency is marked by red arrows for data at 2.5 T for illustration purposes. For the 10 QL sample, above 5 T, the inflection point in the real part of the Faraday rotation moves above our high frequency range and the low-frequency tails becomes flat and overlaps even higher field data.  In our TDTS measurements, top and bottom states are measured simultaneously and so the quantized Faraday rotation is given by Eq. \ref{eq3}.

Since the resolution of our THz  polarimetry is within 1 mrad, we conclude that 10 QL sample enters the quantized regime when field is above 5.75 T with its low-frequency tails falling on the expectation for the  $5  \frac{2\alpha}{1+n}$ plateau. Similarly, as shown in Fig. \ref{Fig2}(e) 6 QL, 8 QL, 12 QL and 16 QL, the low-frequency Faraday rotations falls on the $2 \frac{2\alpha}{1+n}$, $4 \frac{2\alpha}{1+n}$, $7  \frac{2\alpha}{1+n}$ and $7  \frac{2\alpha}{1+n}$ plateaus respectively.  Aside from the filling factor differences, the only qualitative differences between samples is that thicker samples have a narrower magnetic field range where the Faraday rotation is quantized because they have a slightly higher carrier density and filling factor at the same magnetic fields (See SI section \textbf{4} ). 

In our experiment, we measure the top and bottom surfaces of the thin film simultaneously.   Essential for our interpretation in terms of  axion electrodynamics is that we can treat the top and bottom surfaces independently.  Previous ARPES work \cite{zhang2010crossover} showed that the hybridization gap from top and bottom surfaces was an exponential function of thickness for film thicknesses less than 6 QL.   Extrapolating this data to thicker films (as done in Ref. \cite{kim2013coherent}) gives a hybridization gap $\Delta$ of 1 meV at 10 QL.  Treating the hybridization between top and bottom surfaces as a simple two level system (and assuming a strictly linear massless Dirac spectrum) with degenerate Dirac points on top and bottom surfaces, shows that the mixing between top and bottom surfaces will be of order $\Delta/4 E_F$.   For our 10 QL sample this is approximately 0.005 which should be considered negligible.   12 QL and 16 QL samples have even smaller hybridization gaps predicted of 0.3 meV and 0.01 meV respectively and hence have even less overlap of the surface states. Note that theoretical calculations of finite-size effects in Bi$_2$Se$_3$ give a systematically even smaller $\Delta$ at a given thickness (with negligible coupling at 6 QL), which if realized would even further suppress any mixing of top and bottom surfaces \cite{Linder09a}.

We can contrast the above quantized response with DC transport that has shown quantum Hall resistivity plateaus in these films only above $\sim$ 24 T as shown for a typical sample 8 QL film in Fig. \ref{Fig2}(f).  When an external magnetic field is applied perpendicular to the films, top and bottom surface states are gapped due to LL formation while the side surfaces parallel to the magnetic field remain gapless because a small in-plane field will cause only a shift of the Dirac points \cite{pershoguba2012spin}.  The DC QHE in conventional 2DEGs is usually regarded as occurring through ballistic 1D chiral states formed at the edge of the sample.  In the present case, we believe the DC QHE is corrupted at low fields by the non-chiral side states as shown in Fig. \ref{Fig3} (b).  Although these side surface states can be gapped by an amount $\hbar  \textbf{v}_F /d$ through finite size effects (where $d$ is the film thickness), in order that they not contribute to DC transport, this gap must be larger than the Landau level spacing   $E = \textbf{v}_F  \sqrt{2 N e \mathbf{B} \hbar }$ \cite{vafek2011quantum}.   This condition is hard to fulfill with  films thick enough to be effectively 3D and with fields large enough to establish a surface QHE.  We believe that quantized DC transport is achieved in high fields, because the non-chiral side states localize in high magnetic field in the highly disordered edges.  In the present experiment, THz radiation is focused onto a local spot far from the edges of the film so irrespective of their properties they cannot contribute to the spectral response.   The Hall response measured here originates in the ``bulk" of the sample (topological surface states here) and the edge state picture does not apply.   Please see SI section \textbf{7} for further discussion on the AC QHE and the additional issue of how the incompressible bulk responds to an oscillating charge density in an AC experiment.

Data in Fig. \ref{Fig2} gives evidence for a Faraday rotation set by the fine structure constant.   However, such measurements by themselves are limited, as Eq. \ref{eq3} shows that the Faraday rotation still depends non-universally on the index of refraction of the substrate $n$ and our ability to measure the fine structure constant to high precision is limited by our knowledge of it.    However, by using the explicit time structure of TDTS we can define and measure a quantity that depends only on the fine structure constant (and surface filling factors).   When THz light is transmitted through a film and substrate, the substrate itself can be used as an optical resonator \cite{ValdesAguilarPRL12,HancockPRL11} resulting in a series of pulses that each have different histories of interaction with the film as shown in Fig. \ref{Fig3}(a).  The 1st peak that is transmitted through the film undergoes a Faraday rotation, whereas the 2nd peak undergoes an additional reflection and Kerr rotation $\phi_K$.  By subtracting the Faraday rotation, we can measure the Kerr rotation separately.   In the quantized regime, one can show (see SI) that the Kerr rotation (up to factors second order in $\alpha$) is

\begin{equation}
\tan(\phi_K)=\frac{4n\alpha}{n^2-1}(N_t+\frac12+N_b+\frac12).
\label{eq4}
\end{equation}

Representative data for the 10 QL sample for the Kerr rotation is shown in Figs. \ref{Fig3}(c)(d).  Similar to the Faraday rotation, the signatures of cyclotron resonance are inflection points in the real part and dips in the imaginary part.  Above 5.75 T, the Kerr rotation of a 10 QL sample is quantized as $5 \frac{4n\alpha}{n^2-1} $ to within  our experimental resolution at frequencies below 0.8 THz.  The prefactor of 5 is the same as arrived at in the Faraday rotation.  We measured Kerr rotation on samples with different thickness, 6 QL, 8 QL, 12 QL and 16 QL and in all cases the rotation is accurately given by $\frac{4n\alpha}{n^2-1}$ times the filling factor found in the Faraday rotation experiments as shown in Fig. \ref{Fig3}(e).  Inspection of Eqs. \ref{eq3} and \ref{eq4} show that by combining Faraday and Kerr rotation one can eliminate the dependence on the index of the substrate and measure the fine structure constant directly. One finds that

 \begin{equation}
\alpha_{measured}=   \frac{1}{N_t+N_b+1/2+1/2}
  \frac{\tan(\phi_F)^2-\tan(\phi_F)\tan(\phi_K)}{\tan(\phi_K)-2 \tan(\phi_F)}.
  \label{alphaeq}
\end{equation}

Measuring these quantities in a single scan and taking ratios in this fashion also serves to minimize the systematic noise in the output for $\alpha_{measured}$.  Using $N_t+N_b+1/2+1/2 = 5$ and $7$ for the 10 QL and 12 QL samples respectively, we plot the results of Eq. \ref{alphaeq} for two samples in Fig. 4 and find for both the measured value is close to 1/137 ($\sim$ 7.3) mrad.  Averaging over the frequency range that quantized rotation is observed ($\textless$ 0.8 THz) for all samples measured, we find a best measured value for $\alpha_{meas}$ of 1/137.9, which is close to the accepted value 1/137.04. This is the first direct measurement of the finite structure constant based on a topological invariant (and in a purely solid state context).  Although the level of precision we have achieved for $\alpha$ is far less than for instance its determination via the anomalous magnetic moment of the electron \cite{Mohr00a}, the quantization should be considered quite good.   Its deviation from the accepted value is approximately 0.5$\%$, which can be compared favorably to the quantization seen in the quantum spin Hall effect which was only quantized to the 10$\%$ level \cite{konig2007quantum}.  We will also point out that the observed quantization is far better than observed previously in the AC QHE of 2DEGs like GaAs heterostructures and graphene \cite{ikebe2010optical,shimano2013quantum}.  If this measurement could be further refined, it could, along with measures of the Josephson effect and quantized Hall resistances in 2DEGs provide a purely solid state measure in a redefined conventional electrical unit scheme for the impedance of free space $Z_0 = \sqrt{\mu_0 / \epsilon_0}$.   Currently, both $\mu_0$ and $\epsilon_0$ are defined quantities in the International System of Units.  With $\mu_0$ staying a defined quantity (and essential for the definition of the Ampere), a measure of $\alpha$ would allow $c$ to become a measured quantity in a condensed matter experiment.

It is important to distinguish our results from a conventional AC QHE as may be observed in a 2DEG.  Do we truly probe `axion electrodynamics' and the TME effect?  As discussed above, the TME is characterized by a $\theta$ angle that is $2 \pi (N+\frac{1}{2})$ or equivalent a half-integer QHE effect.  In Fig. \ref{Fig4} (c), we plot the observed quantization index versus the total filling factor (which can be measured independently as discussed above and in the SI section \textbf{4, 6}).   One can see that there is a systematic offset of `1' in the position of the plateaus that originates from the Berry's phase.   With our previous results establishing surface state transport from two independent surfaces, one must associate a contribution to this offset of 1/2  for each surface by itself.  This establishes a $\pi$ value of the axion angle of the topological insulators and the TME.   Note that our experimental configuration does not allow us to establish a net magnetic/electric moment of the whole sample by external electric/magnetic field which would also be evidence for the TME.  This would require creating  half-integer QHE effects on opposing surfaces of opposite sign (as compared to the same sign half-integer QHE in the current experiment.   However, as explained above, the present effect arises from the same $\theta \textbf{E} \cdot \textbf{B}$ ($\theta=\pi$ here) and is a probe of equivalent physics. Note that in materials like Cr$_2$O$_3$, low point group symmetry ensures the existence of the magnetoelectric effect but not the magnitude of tensor elements.  In TIs, time-reversal or inversion symmetry ensure that the magnetoelectric response (the $\theta$ angle) is quantized.

With advances in high-quality Bi$_2$Se$_3$ films and charge-transfer doping by MoO$_3$ to lower the chemical potential to as low as $\sim$ 30 meV above the Dirac point coupled to the development of high-precision time-domain THz polarimetry, we have observed quantized Faraday rotation and Kerr rotation from topological surface states.  The quantized responses we find here should not be viewed as a simple manifestation of the quantized quantum Hall transport seen in usual 2DEGs, since TI surface states live on a closed 2D manifold embedded in 3D space.  For instance, our DC transport does not show a quantized Hall resistance in this field range.  In a formalism where the TIs are described as bulk magnetoelectrics, this response can be described in the context of a topological magnetoelectric effect and axion electrodynamics.  Among other things, this provides a direct solid-state measure of the fine structure constant.   The observation of this effect gives a definitive characterization of topologically insulating states of matter.   It may prove to be an essential tool in the discovery of theoretically anticipated states of matter such as fractional topological insulators in the form of a fractional magnetoelectric effect \cite{maciejko2010fractional, Swingle11a}.
 
We would like to thank M. Franz, T. Hughes, A. MacDonald, J. Maciejko, J. Moore, M. Orlita, V. Oganesyan, W.-K. Tse, A. Turner, R. Vald\'es Aguilar, X. L. Qi and S.C. Zhang for helpful discussions.  Experiments were supported by the Army Research Office Grant W911NF-15-1-0560 with additional support by the Gordon and Betty Moore Foundation through Grant GBMF2628 to NPA at JHU.  Film growth for this work was supported by the NSF DMR-1308142, EFMA-1542798 and the Gordon and Betty Moore Foundation EPiQS Initiative Grant GBMF4418 to SO at Rutgers.

Note added: we noticed two other studies\cite{okada2016observation, dziom2016observation} on the same subject that were posted on arXiv around a similar time. Different from our work, Ref. \citeonline{okada2016observation} is towards quantization but not yet quantized  and Ref. \citeonline{dziom2016observation} has a gate. 

\bibliography{TopoIns}

\clearpage

\begin{figure}[htp]
\includegraphics[width=0.5\columnwidth]{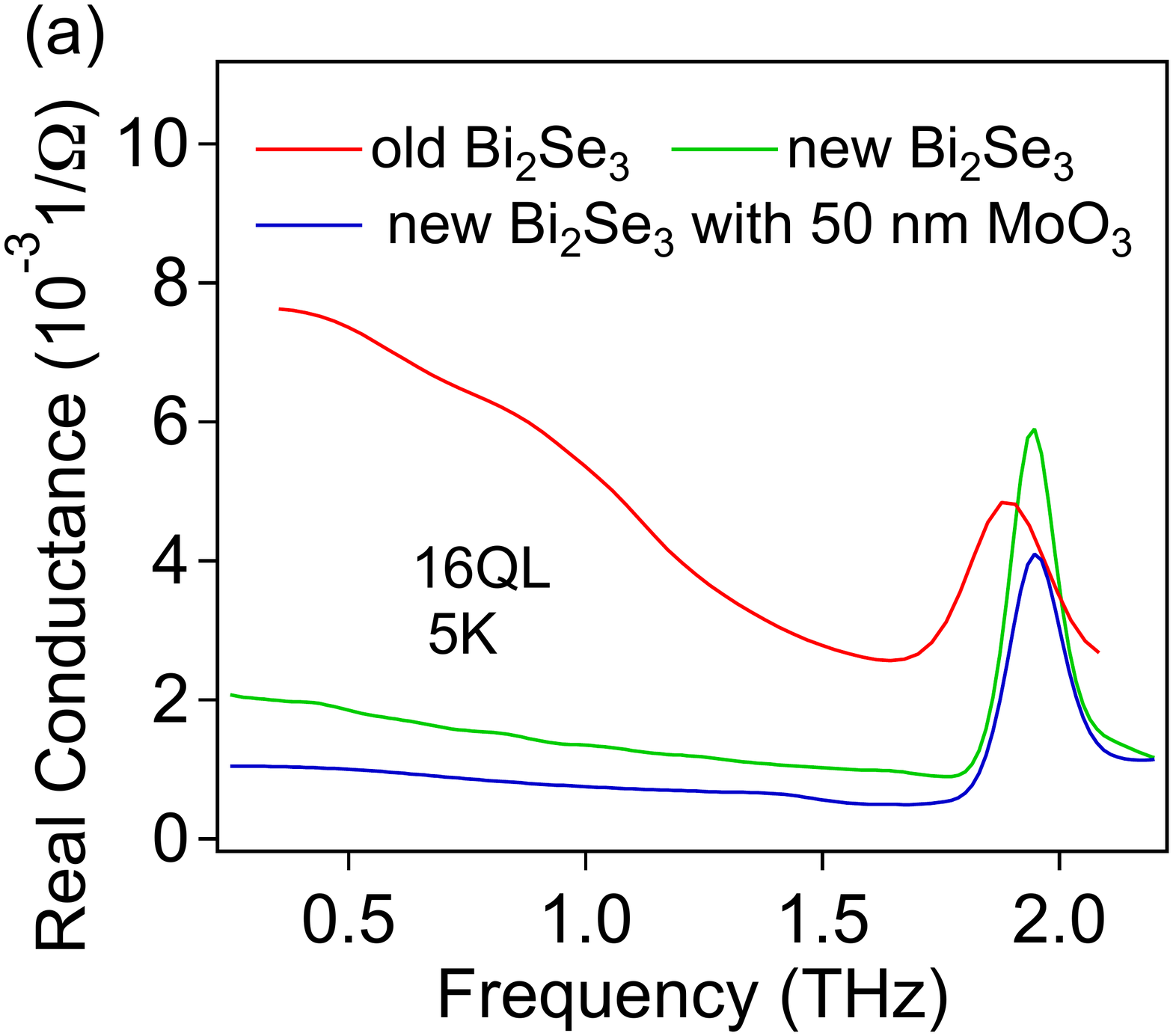}
\includegraphics[width=0.5\columnwidth]{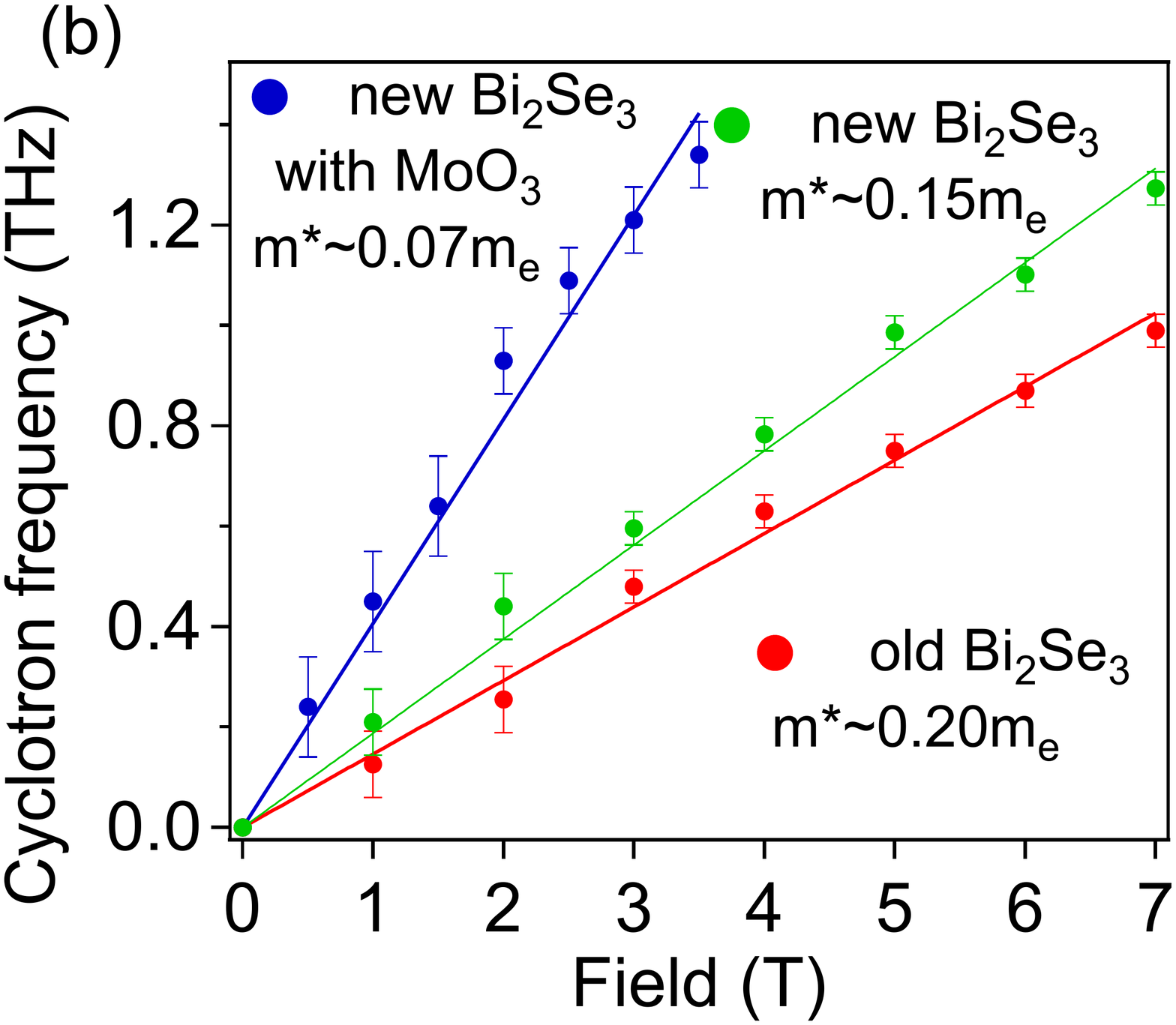}

\caption{(Color online) (a) Zero-field real conductance (b) Cyclotron frequencies ($\omega_c/2\pi$) vs. field of 16 QL old Bi$_2$Se$_3$, new Bi$_2$Se$_3$ and new Bi$_2$Se$_3$ with MoO$_3$ at 5 K.  The solid lines in (b) are linear fits that give the effective mass from $\frac{\omega_c}{2\pi} = \frac{e B}{m^*}$  in the low field classical cyclotron resonance regime.  }
 \label{Fig1}
\end{figure} 

\begin{figure*}[htp]
\includegraphics[width=0.5\columnwidth]{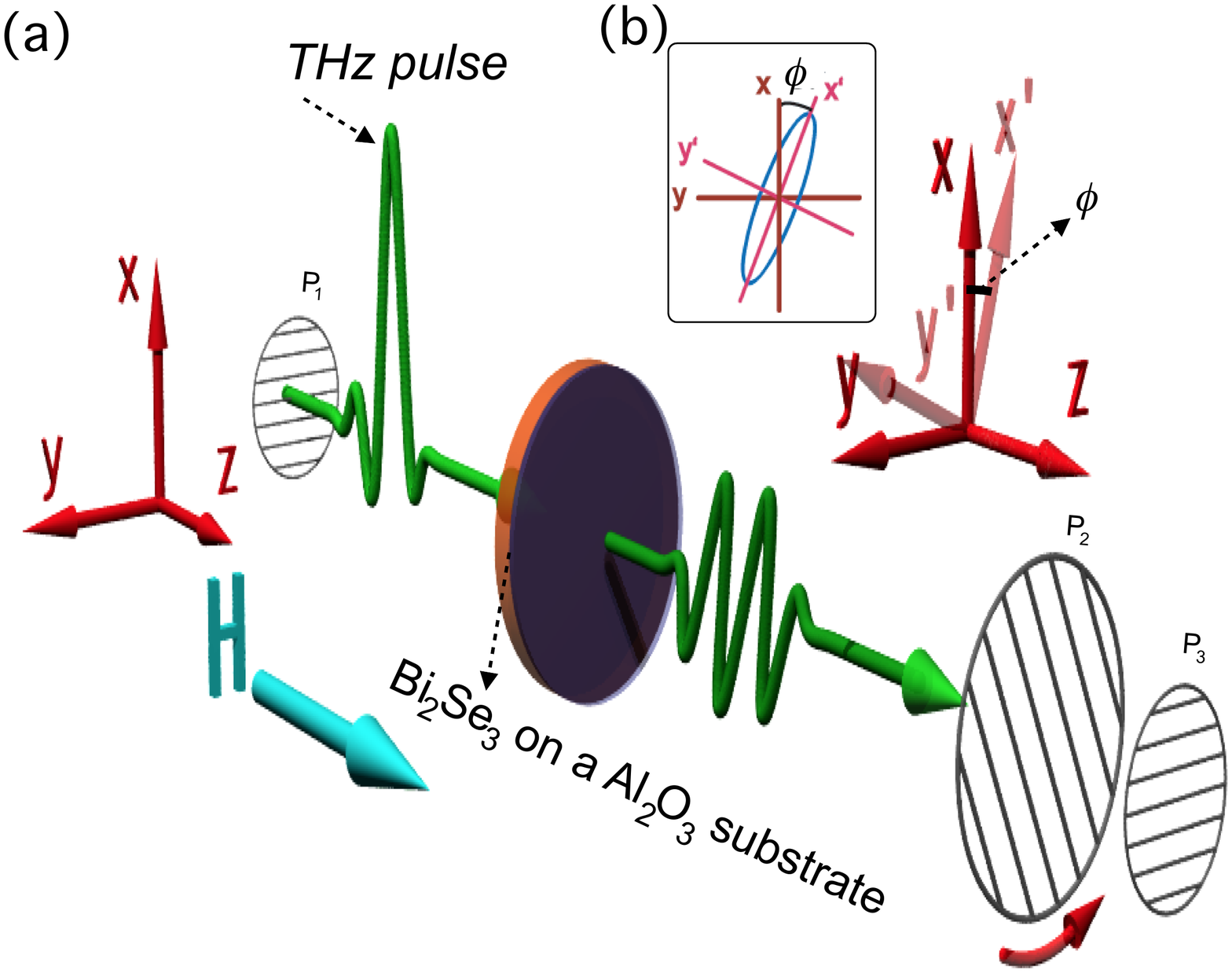}
\includegraphics[width=0.5\columnwidth]{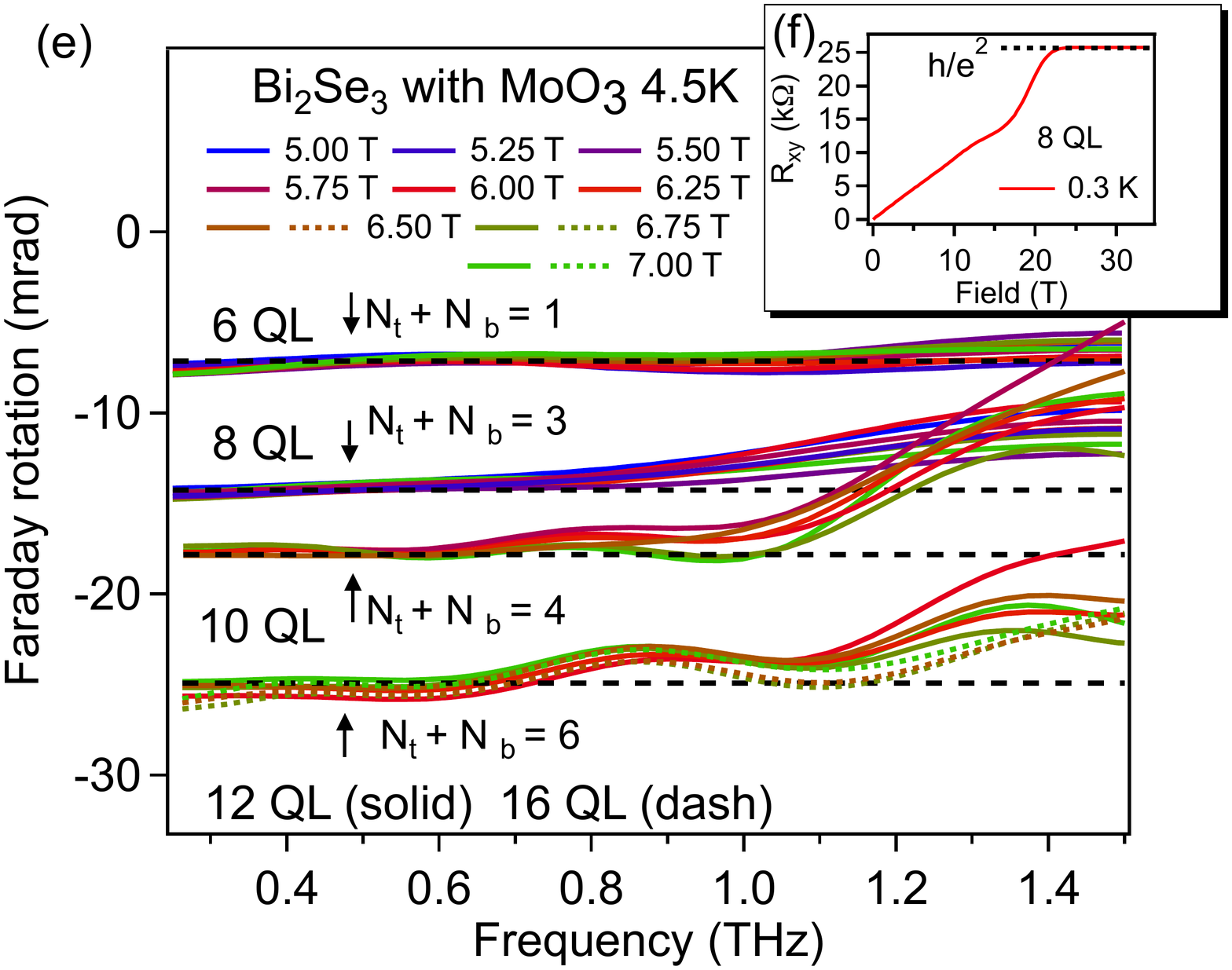}
\includegraphics[width=0.5\columnwidth]{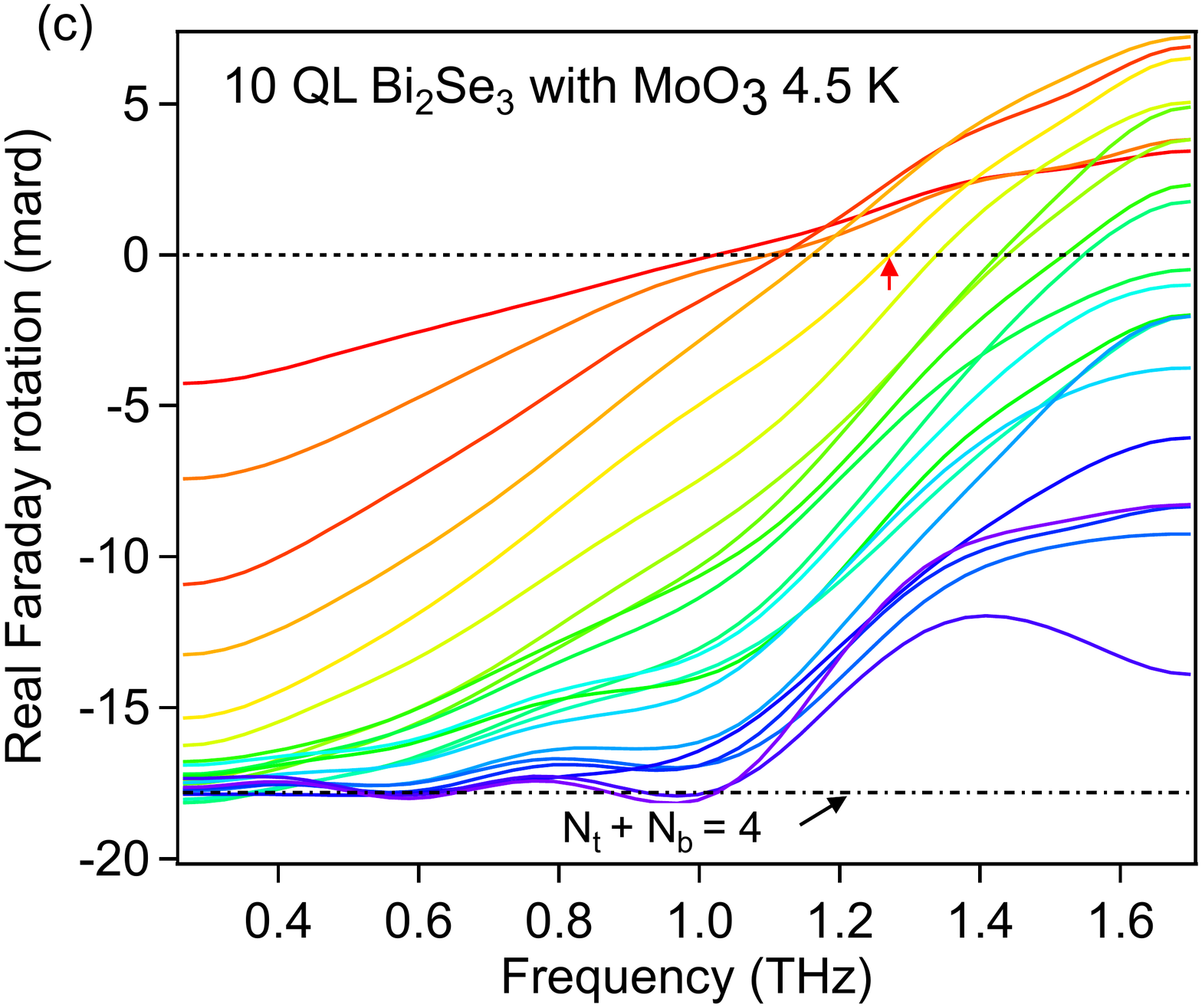}
\includegraphics[width=0.5\columnwidth]{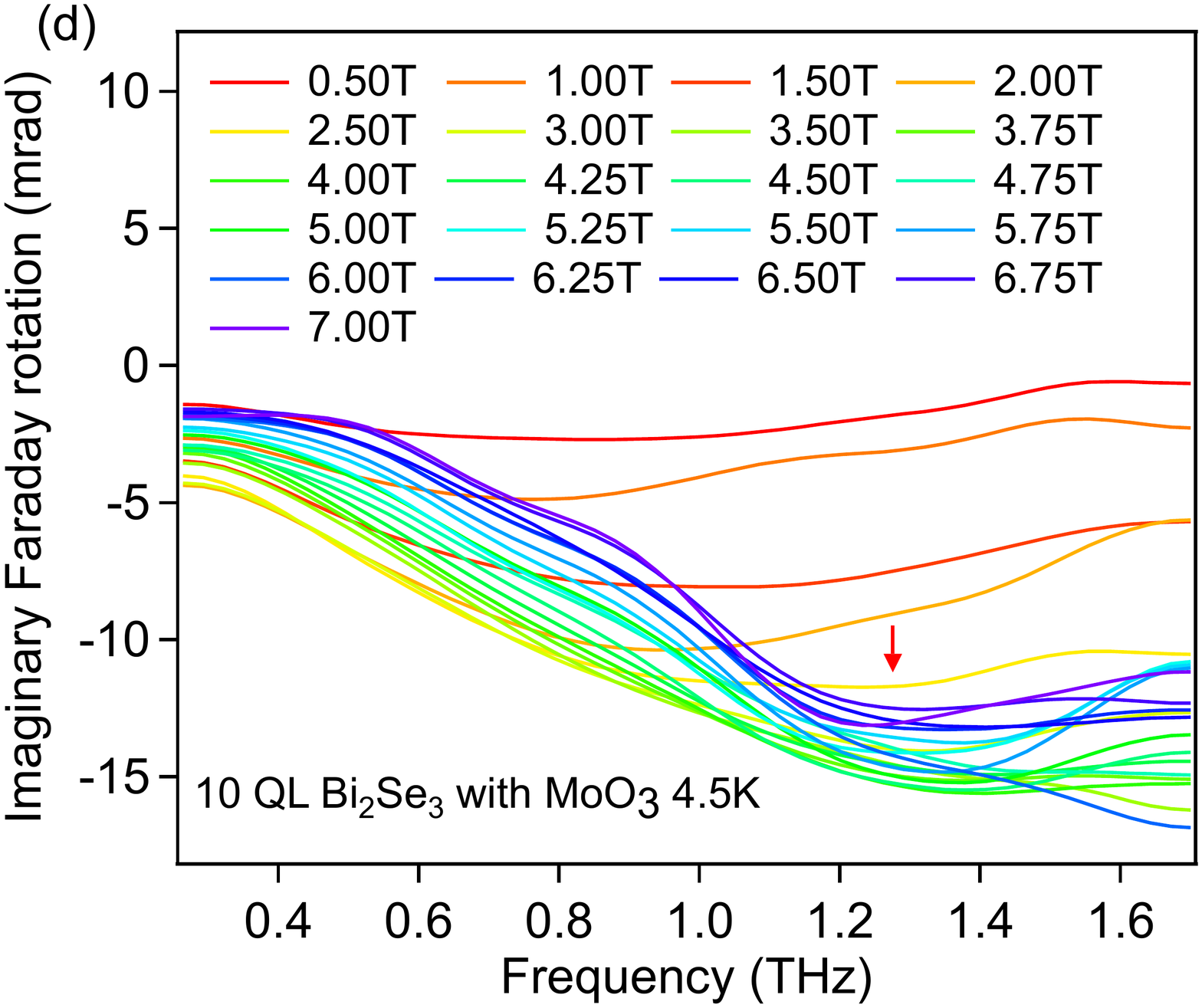}

\caption{(Color online) (a) Schematic diagram of the Faraday rotation experiment. P$_1$, P$_2$ and P$_3$ are polarizers. The polarization plane ($xz$) of the linearly polarized incoming THz beam is rotated by the Faraday angle $\phi_F$ (into the $x'z$ plane) after passing through Bi$_2$Se$_3$ on a sapphire substrate in a perpendicular magnetic field ($z$ direction). The polarization acquires an ellipticity simultaneously, as shown in (b). (c) Real  part of Faraday rotation of 10 QL new Bi$_2$Se$_3$ with MoO$_3$ at 4.5 K.   The dashed line is the expectation from Eq. \ref{eq3}.  (d) Imaginary part of Faraday rotation.  (e) Quantized Faraday rotation for different samples. Dashed lines are theoretical expectation values assuming certain values for the filling factor of the surface states. (f) DC transport Hall resistance of a representative 8 QL sample.}
 \label{Fig2}
\end{figure*} 

\begin{figure*}[htp]
\includegraphics[width=0.5\columnwidth]{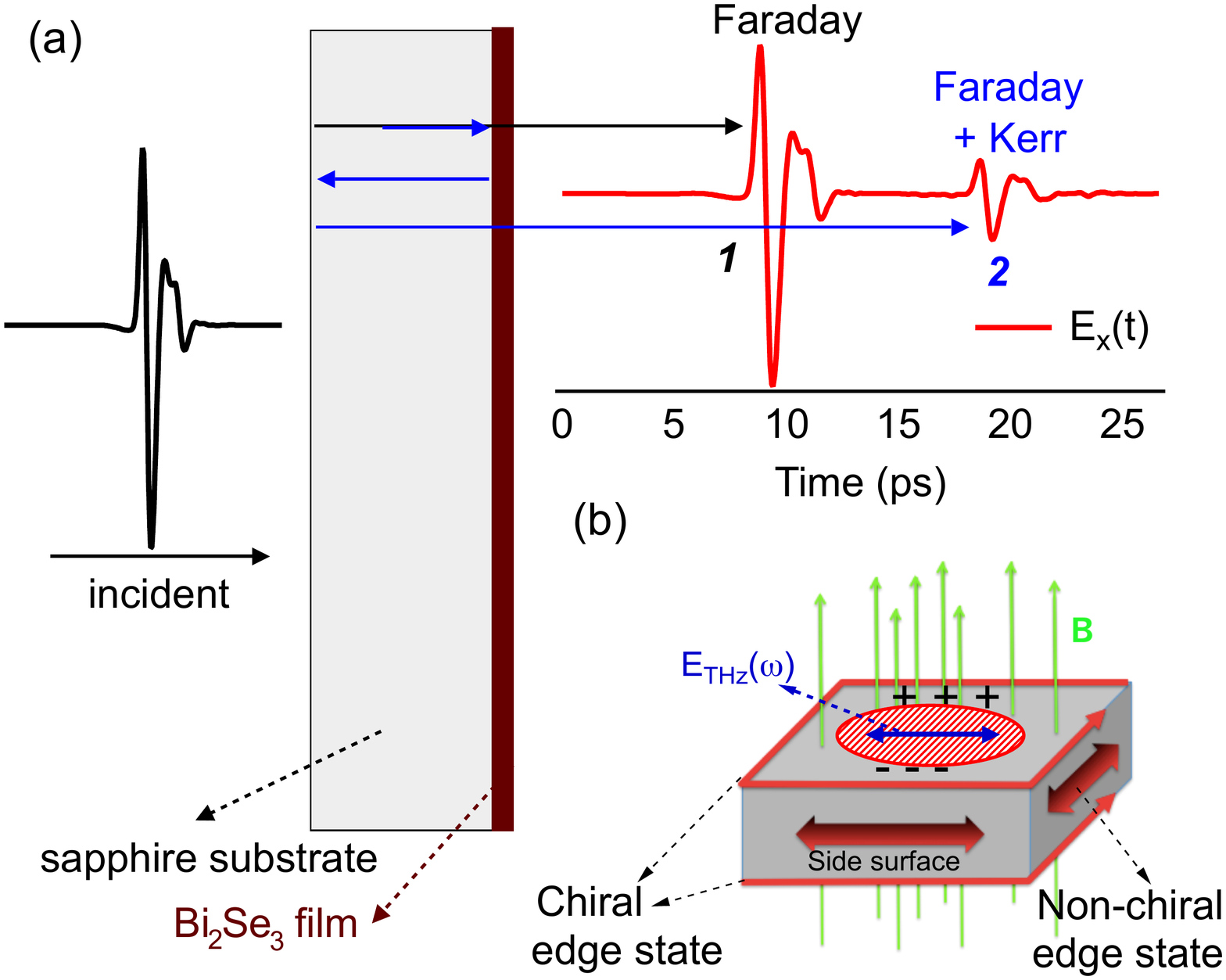}
\includegraphics[width=0.5\columnwidth]{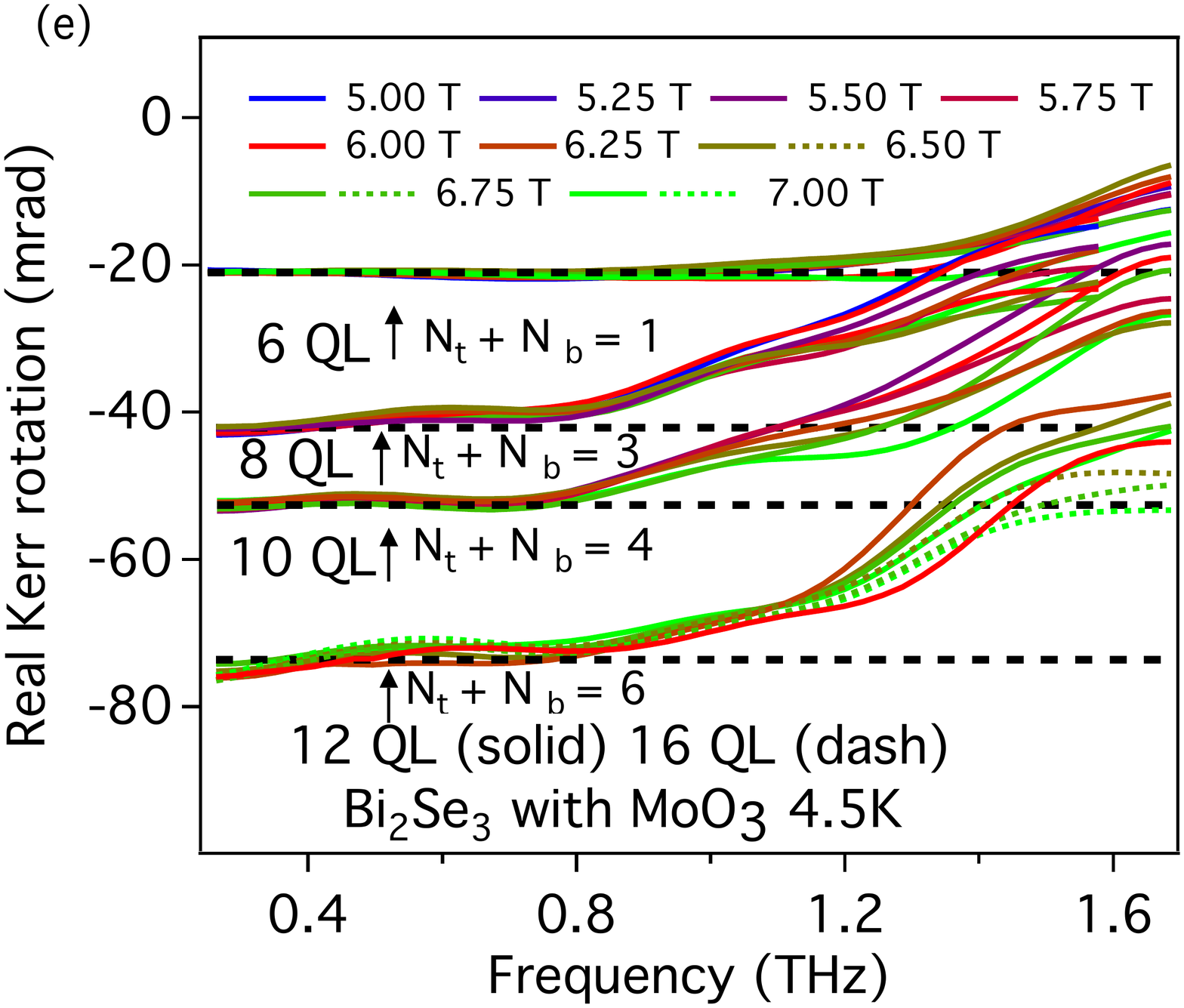}
\includegraphics[width=0.5\columnwidth]{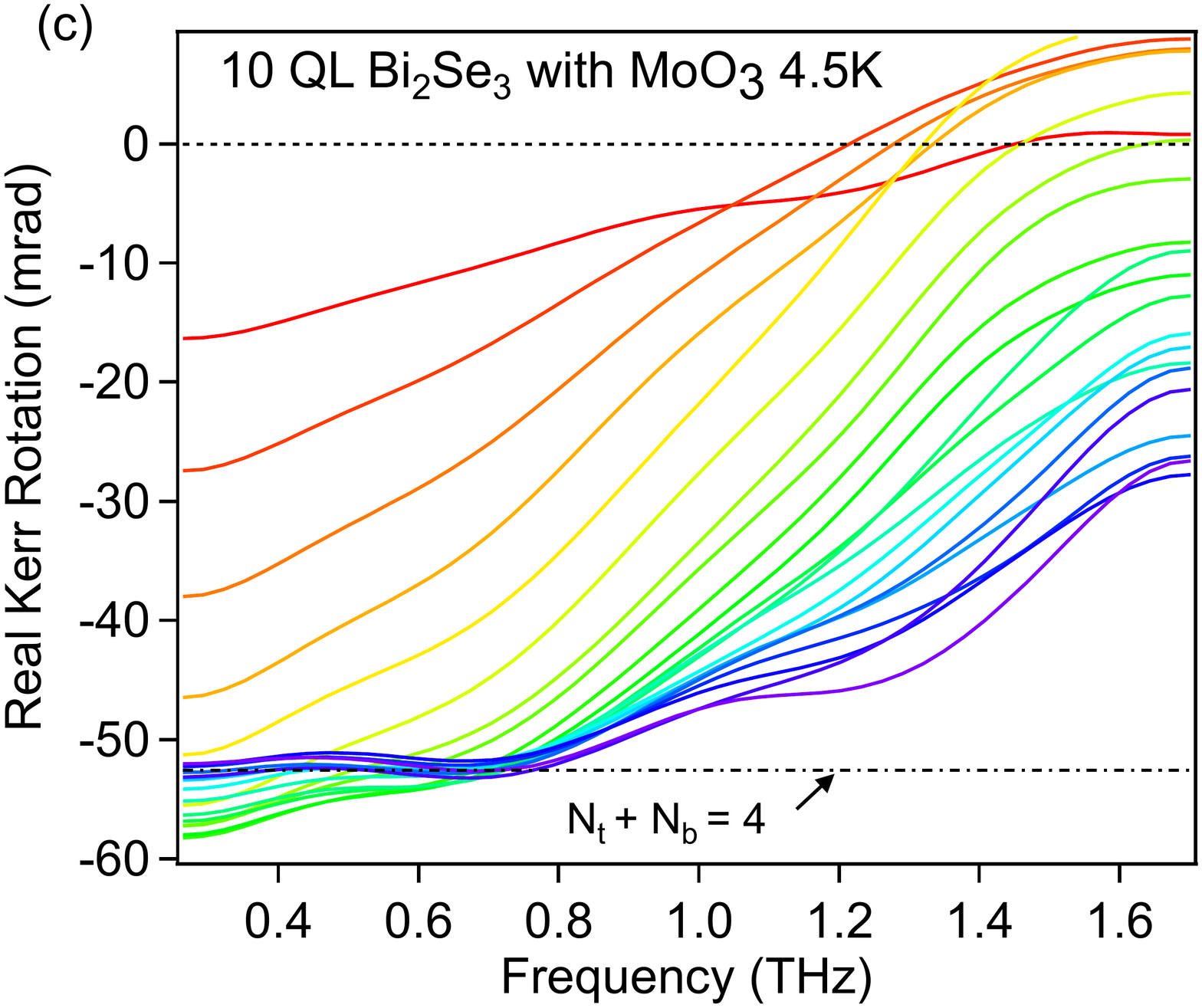}
\includegraphics[width=0.5\columnwidth]{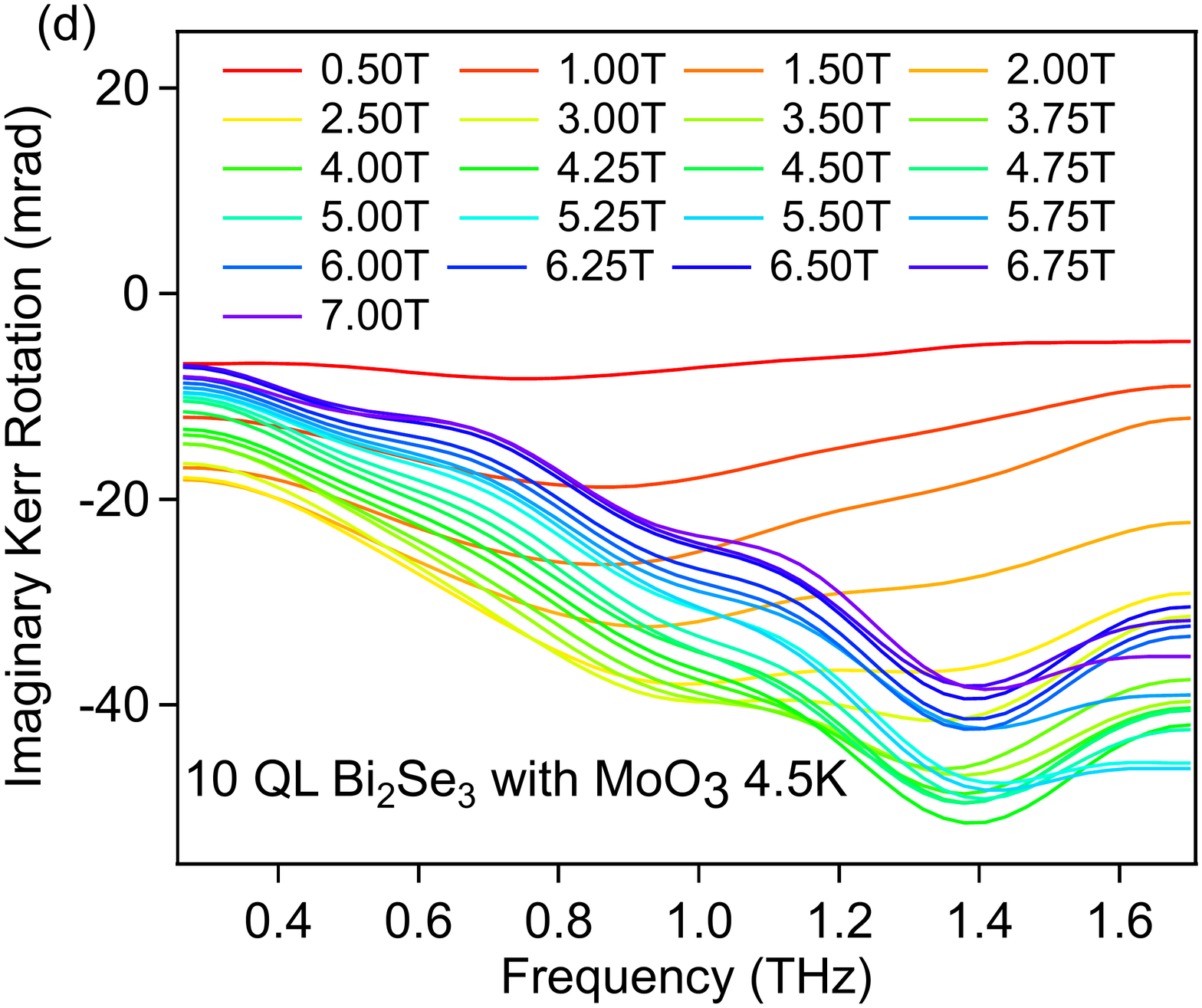}
\caption{(Color online) (a) Schematic diagram of the Kerr rotation experiment. The black and blue arrows shows the optical path for the 1st and 2nd pulses in the time trace. (b)  Bi$_2$Se$_3$ thin film in a magnetic field (substrate not shown). Nonchiral edge states from the side surface states are shown by an two-direction arrow. The circle on the top surface indicates the THz spot where transverse oscillating charge density is built up by the THz electric field. (c) Real  (d) Imaginary part of Kerr rotation of 10 QL new Bi$_2$Se$3$ with MoO$_3$ at 4.5 K. (e) Quantized Kerr rotation for different samples.}
 \label{Fig3}
\end{figure*} 

 \begin{figure}[htp]
\includegraphics[width=0.5\columnwidth]{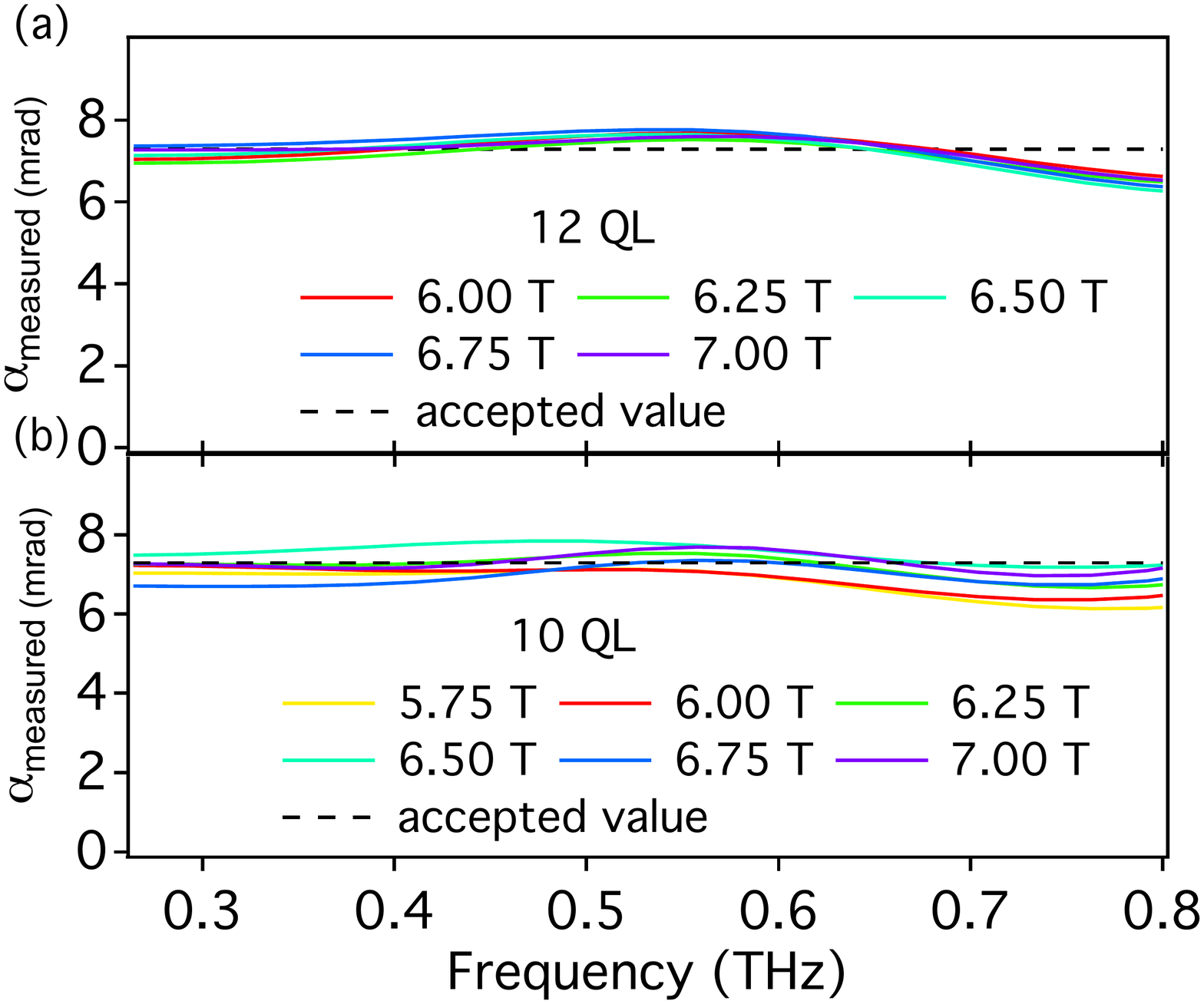}
\includegraphics[width=0.5\columnwidth]{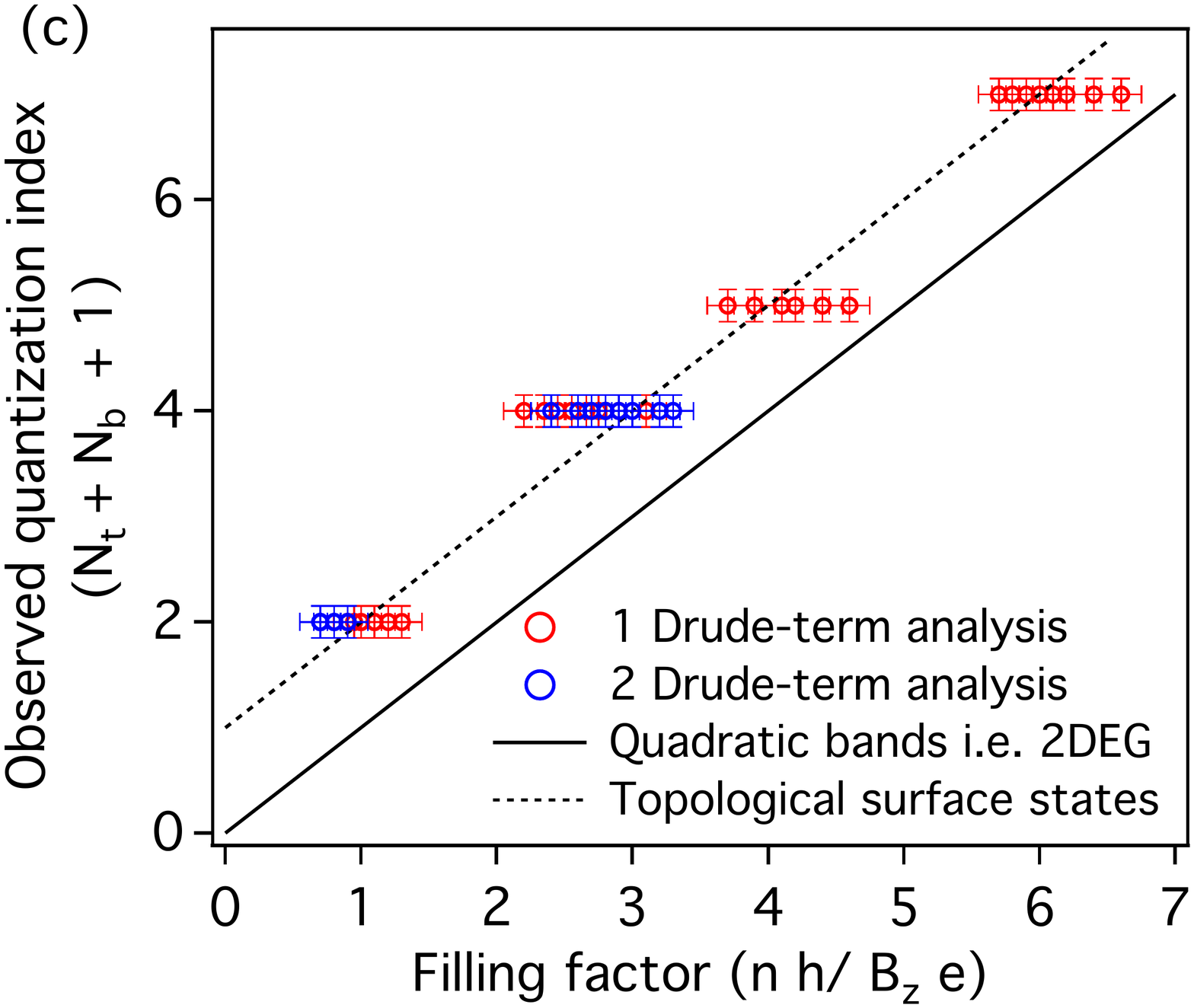}
\caption{(Color online) The measured fine structure constant of (a) 12 QL (b) 10 QL  new Bi$_2$Se$_3$ with MoO$_3$ at different field at 4.5 K. (c) Measured quantization index vs. filling factor. The solid line is the expectation for quadratic bands and the dashed line is for two topological surface states.  }
 \label{Fig4}
\end{figure}

\clearpage
\newpage
\textbf{\Large Supplementary Information for 
``Quantized Faraday and Kerr rotation and axion electrodynamics of the surface states of three-dimensional topological insulators''} 

\bigskip

\section{Derivation of Modified Maxwell's Equations}

Qi et al. \cite{Qi08b} showed that the electrodynamics of topological insulators can be described by adding an additional topological term $ \mathcal{L}_{\theta} = - 2 \alpha \sqrt {\frac{ \epsilon_0}{\mu_0}}  \frac{\theta}{2 \pi}  \mathbf{E} \cdot  \mathbf{B}  $ to the usual Maxwell Lagrangian $\mathcal{L}_0$.   Here we rederive the modified Maxwell's equations in the conventional 3D vector component notation, which will be more familiar to many readers of this section as compared to the relativistic Einstein notation that is typical in the field theory literature.  The two terms in the  Lagrangian density written down in potential form are

\begin{eqnarray}
\mathcal{L}_0 =  \frac{  \epsilon_0 }{2} ( \nabla  \phi + \frac{\partial \mathbf{A} }{  \partial t})^2  - \frac{1}{2 \mu_0}   ( \nabla \times  \mathbf{A})^2  -   
\rho \phi +  \mathbf{J} \cdot   \mathbf{A} ,  \\
\mathcal{L}_{\theta} = - 2 \alpha \sqrt{\frac{ \epsilon_0 }{\mu_0} } \frac{ \theta}{ 2 \pi}  ( \nabla  \phi + \frac{\partial \mathbf{A} }{  \partial t} )\cdot ( \nabla \times  \mathbf{A}) .
\label{Lagrang}
\end{eqnarray}

Here $\alpha$ is the fine structure constant and $\epsilon_0$ and $\mu_0$ are the  permittivity and permeability of the free space.  The topological contribution $\theta$ is an angle $2 \pi (N + \frac{1}{2} )$ where $N$ is an integer that indicates the highest fully filled Landau level (LL) of the surface when TRS is broken weakly.  As suggested by its form and this definition, it can be formulated as a bulk quantity only modulo a quantum (here $2 \pi$) in much the same way as the electric polarization  \textbf{P} in a ferroelectric can only be defined as a bulk quantity modulo a dipole quantum that depends on the surface charge \cite{KingSmith93a,Essin09a}.  The action is 

\begin{equation}
\mathcal{S} =  \mathcal{S}_0 + \mathcal{S}_\theta =   \int dt \: d^3 \! x \:   ( \mathcal{L}_0  + \mathcal{L}_{\theta})
\label{Action}
\end{equation}

\noindent where $ \mathcal{S}_\theta $ derives from the additional term and $ \mathcal{S}_0 $ is the usual Maxwell action.  One can start with with Eq. \ref{Action} and perform the typical variation of the potentials in the action to get modifications to Gauss's law and Amp\`ere's law.   The modified Gauss's law term comes from variations in the scalar potential  $\phi$.   One defines

\begin{equation}
\delta \mathcal{S} =   \mathcal{S}(\phi + \delta \phi)  -  \mathcal{S}(\phi)  = \delta  \mathcal{S}_0 +  \delta  \mathcal{S}_\theta
\label{delta}
\end{equation}

\noindent where $ \delta \phi$ is an infinitesimal.  As found in standard references \cite{peskin1995introduction} the Maxwell part of the variation can be written as

\begin{equation}
\delta  \mathcal{S}_0 = -  \int dt \: d^3 \! x \: [   \epsilon_0 \nabla \cdot (  \nabla  \phi +  \frac{ \partial \mathbf{A}}{  \partial t} ) + \rho ] \delta \phi.
\end{equation}

For the new term, to first order in $ \delta \phi$ one has the variation

\begin{equation}
\delta  \mathcal{S}_\theta =  - \int dt \: d^3 \! x \:  [    2 \alpha \sqrt{\frac{ \epsilon_0 }{\mu_0} } \frac{ \theta}{ 2 \pi}  ( \nabla \times  \mathbf{A}  ) \cdot \nabla \delta \phi  ].
\end{equation}

As with the Maxwell term, one shifts the derivatives to the other spatially dependent terms in the integrand by integration by parts.  The surface terms can be set to zero.  One has 

\begin{equation}
\delta  \mathcal{S}_\theta =  \int dt \: d^3 \! x \:   \nabla \cdot [   2 \alpha \sqrt{\frac{ \epsilon_0 }{\mu_0} } \frac{ \theta}{ 2 \pi}  ( \nabla \times  \mathbf{A}  )  ] \delta \phi.
\end{equation}

Expanding the divergence and using the fact that the divergence of a curl is zero one has

\begin{equation}
\delta  \mathcal{S}_\theta =  \int dt \: d^3 \! x \:   [ \nabla ( \frac{ \theta}{2 \pi} ) \cdot    2 \alpha \sqrt{\frac{ \epsilon_0 }{\mu_0} }  ( \nabla \times  \mathbf{A}  )  ] \delta \phi.
\end{equation}

We add this to the variation of the usual Maxwell action to get 

\begin{equation}
\delta  \mathcal{S} =   \int dt \: d^3 \! x \:  [ - (  \epsilon_0 \nabla \cdot (  \nabla  \phi +  \frac{ \partial \mathbf{A}}{  \partial t} ) + \rho )  + 2 \alpha \sqrt{\frac{ \epsilon_0 }{\mu_0} }   \nabla ( \frac{ \theta}{2 \pi} )   \cdot ( \nabla \times  \mathbf{A}  )  ] \delta \phi.
\end{equation}

Setting the variation of this total action to zero requires the term in the brackets above be equal to zero.  Rearranging and substituting back in for the fields, one gets a modified version of Gauss's law with an additional source term that reads

\begin{equation}
\nabla \cdot   \mathbf{E} = \frac{\rho}{\epsilon_0} -  2 c \alpha  \nabla (\frac{\theta}{2 \pi}) \cdot   \mathbf{B}.
\label{Gauss}
\end{equation}

To get the modified version of  Amp\`ere's law one must vary the vector potential.  Expanding $ \mathcal{S}( \mathbf{A} +  \delta \mathbf{A}     )$ to first order in $ \delta \mathbf{A}  $ one has for the Maxwell term

\begin{equation}
\delta  \mathcal{S}_0 =  \int dt \: d^3 \! x \:  [ - \epsilon_0 \frac{\partial (\nabla \phi + \partial \mathbf{A}/ \partial t )}{ \partial t} - \nabla \times (\nabla \times  \mathbf{A} ) / \mu_0 + \mathbf{J}]  \cdot   \delta \mathbf{A}  .
\end{equation}

For $\mathcal{S}_\theta$ we have

\begin{equation}
\delta  \mathcal{S}_\theta =  \int dt \: d^3 \! x \:  [  - 2 \alpha \sqrt{  \frac{\epsilon_0}{ \mu_0} }  \frac{\theta}{ 2 \pi}   (\frac{\partial \delta \mathbf{A} }{\partial t} \cdot (\nabla \times  \mathbf{A}  )   + (\nabla \phi + \frac{\delta  \mathbf{A} }{ \delta t} ) \cdot (\nabla \times \delta  \mathbf{A})  ) ]
\end{equation}

Now we integrate by parts by moving the derivative with respect to time on the first term and the gradient on the second.  Setting  the surface terms to zero and after some simplification one gets

\begin{equation}
\delta  \mathcal{S}_\theta =  \int dt \: d^3 \! x \:  [   2 \alpha \sqrt{  \frac{\epsilon_0}{ \mu_0} }     (   \frac{ \partial \theta/ \partial t}{ 2 \pi}  ( \nabla \times   \mathbf{A} )   - \nabla ( \frac{\theta}{ 2 \pi}) \times ( \nabla \phi + \frac{\partial \mathbf{A} }{ \partial t}  ) ] \cdot \delta \mathbf{A} .
\end{equation}

The total variation with respect to the vector potential then reads  

\begin{multline}
\delta  \mathcal{S} =  \int dt \: d^3 \! x \:  [ - \epsilon_0 \frac{\partial (\nabla \phi + \partial \mathbf{A}/ \partial t )}{ \partial t} - \nabla \times (\nabla \times  \mathbf{A} ) / \mu_0 + \mathbf{J}   \\ +  2 \alpha \sqrt{  \frac{\epsilon_0}{ \mu_0} }     (  \frac{ \partial \theta/ \partial t}{ 2 \pi}   ( \nabla \times   \mathbf{A} )   - \nabla ( \frac{\theta}{ 2 \pi}) \times ( \nabla \phi + \frac{\partial \mathbf{A} }{ \partial t}  ) )     ]  \cdot   \delta \mathbf{A}  .
\end{multline}

As before if the total variation is to be zero for any infinitesimal $ \delta \mathbf{A}$ then the quantity in brackets must be zero.   Rearranging and again substituting in for the fields, one finds a modified version of Amp\`ere's law with an additional current term that reads

\begin{equation}
\nabla \times    \mathbf{B} = \mu_0   \mathbf{J}  + \frac{1}{c^2}  \frac{\partial   \mathbf{E}}{\partial    t} +  \frac{2 \alpha}{c} [  \mathbf{B}   \frac{\partial }{\partial    t}   (  \frac{\theta}{2 \pi} )  + \nabla (\frac{\theta}{2 \pi}) \times    \mathbf{E}  ].
\label{Ampere}
\end{equation}

\section{Derivation of Modified Fresnel and Transmission Equations}

One can use the modified Amp\`ere's law Eq. \ref{Ampere} with usual Faraday's law to derive to find modified boundary conditions for the electric and magnetic field and from there reflection and transmission coefficients for a traveling wave.   A non-topological surface current $\mathbf{J}$ could be included in this derivation, but we neglect it here because the chemical potential of our samples are in the gap.  One can show for instance that the modified inverse transmission matrix for transmission across the vacuum-TI interface is

\begin{equation}
\hat{T}^{-1} = 
\left[\begin{array}{cc} \frac{1}{2}(1 + c/v) & \frac{2 \alpha \theta}{2 \pi} \\  - \frac{2 \alpha \theta}{2 \pi}  & \frac{1}{2}(1 + c/v)  \end{array}\right],
\label{ABCD}
\end{equation}

\noindent where $v$ is the speed of light inside the TI.  For a slab system with broken TRS, the transmission is most easily computed with circular eigenstates\cite{armitage2014constraints}.  For incoming light polarized polarized along the $x$ direction, for an isotropic sample the Faraday rotation is given by 
 
\begin{equation}
\tan(\phi_F) = \mathrm{T}_{xy} / \mathrm{T}_{xx} = \frac{1}{i}  \frac{ \mathrm{T}_{++}  -   \mathrm{T}_{--}  }{ \mathrm{T}_{++}  +  \mathrm{T}_{--}},
\label{MatrixForm}
\end{equation}

\noindent where for instance $ \mathrm{T}_{xx}$ and $ \mathrm{T}_{xy}$ are diagonal and off-diagonal terms in the Jones transmission matrix (not to be confused with the Fresnel transmission coefficient for light across an interface given by the inverse of Eq. \ref{ABCD}) in the $xy$ coordinate frame and  $\mathrm{T}_{++}$ and  $\mathrm{T}_{--}$ are diagonal terms in the transmission matrix for the circular basis.  (See Ref. \cite{armitage2014constraints} for details about the various bases of Jones matrices and their introconversions).  In the limit where the thickness of the film is much less than the wavelength of the incident light, the effect of the two conducting surfaces can be treated as single surface with effective conductance $G^{ \mathrm{eff} }= G_t + G_b$.    The transmission then of a thin film on a substrate with index of refraction $n$  referenced to a nominally identical substrate is

\begin{equation}
 \mathrm{T}_{++}  = \frac{1 + n}{ 1 + n + Z_0 G^{ \mathrm{eff} }_{++}} ,
 \label{TransmissionFunction}
 \end{equation}

\noindent were $Z_0 = \sqrt{ \frac{\mu_0}{\epsilon_0}}$ is the impedance of free space.  Small differences in substrate thickness $\Delta L$ from the reference can be easily dealt with by multiplying by a phase factor $e^{i \omega \Delta L (n - 1 ) /c }$.  Eq. \ref{TransmissionFunction} applied to Eq. \ref{MatrixForm} yields

\begin{equation}
\tan(\phi_F) = \frac{Z_0 G^{ \mathrm{eff} }_{xy}} {2 + 2n + Z_0 G^{ \mathrm{eff} }_{xx}},
 \label{FaradayRotation}
 \end{equation}

\noindent in the quantum regime for the TIs $G^{ \mathrm{eff} }_{xx} \rightarrow 0$ and $G^{ \mathrm{eff} }_{xy} = \frac{e^2}{h}  (N_t + \frac{1}{2} + N_b + \frac{1}{2})  $ and 

\begin{equation}
\tan(\phi_F)=\frac{2\alpha}{1+n}(N_t+\frac12+N_b+\frac12),
\label{SIeq3}
\end{equation}

\noindent used in the main text follows.  The expression for $\tan(\phi_K)$ derived in the main text can be derived in a similar fashion.

\begin{equation}
\
\tan{\phi_K}=\frac{2Z_0nG^{eff}_{xy}}{n^2-[1+2Z_0G^{ \mathrm{eff} }_{xx}+Z^2_0(G^{ \mathrm{eff} 2 }_{xx}+G^{ \mathrm{eff} 2 }_{xy})]}.
\label{EqaKerr}
\end{equation}

\noindent In the quantum regime, 

\begin{equation}
\tan(\phi_K)=\frac{4n\alpha}{n^2-1}(N_t+\frac12+N_b+\frac12).
\label{SIeq4}
\end{equation}

Note that this expression neglects a correction coming from the $G^{ \mathrm{eff} 2 }_{xy}$ in the denominator of Eq. \ref{EqaKerr}, that to lowest order goes as $\alpha^3$.  This correction is completely undetectable at the current level of precision of the experiment, but may be important formally. If we consider the correction of $\alpha^3$, 

\begin{equation}
\tan(\phi_K)=\frac{4n\alpha}{n^2-1-4(N_t+\frac12+N_b+\frac12)^2\alpha^2}(N_t+\frac12+N_b+\frac12).
\label{eqaKerr2}
\end{equation}

\noindent we can still write down a universal expression of the fine structure constant in terms of Faraday rotation and Kerr rotation as following and define the topological invariant.

  \begin{equation}
\alpha_{measured}=   \frac{1}{N_t+N_b+1/2+1/2}
  \frac{\tan(\phi_F)^2-\tan(\phi_F)\tan(\phi_K)}{\tan(\phi_K)-2 \tan(\phi_F)-\tan(\phi_F)^2 \tan(\phi_K)}.
  \label{SIalphaeq}
\end{equation}

\noindent Due to the small size of the rotations, the third term in the denominator makes a negligible contribution to the fine structure constant we measure and so we have neglected it in the main text. eThe accuracy of the current experiments is limited by the accuracy of the polarizers in the THz regime. In the future, when the experiment accuracy can be compared with DC quantum Hall effect, it is essential to use Eq. \ref{SIalphaeq} to calculate the fine structure constant.

\begin{figure*}[htp]
\includegraphics[width=0.5\columnwidth]{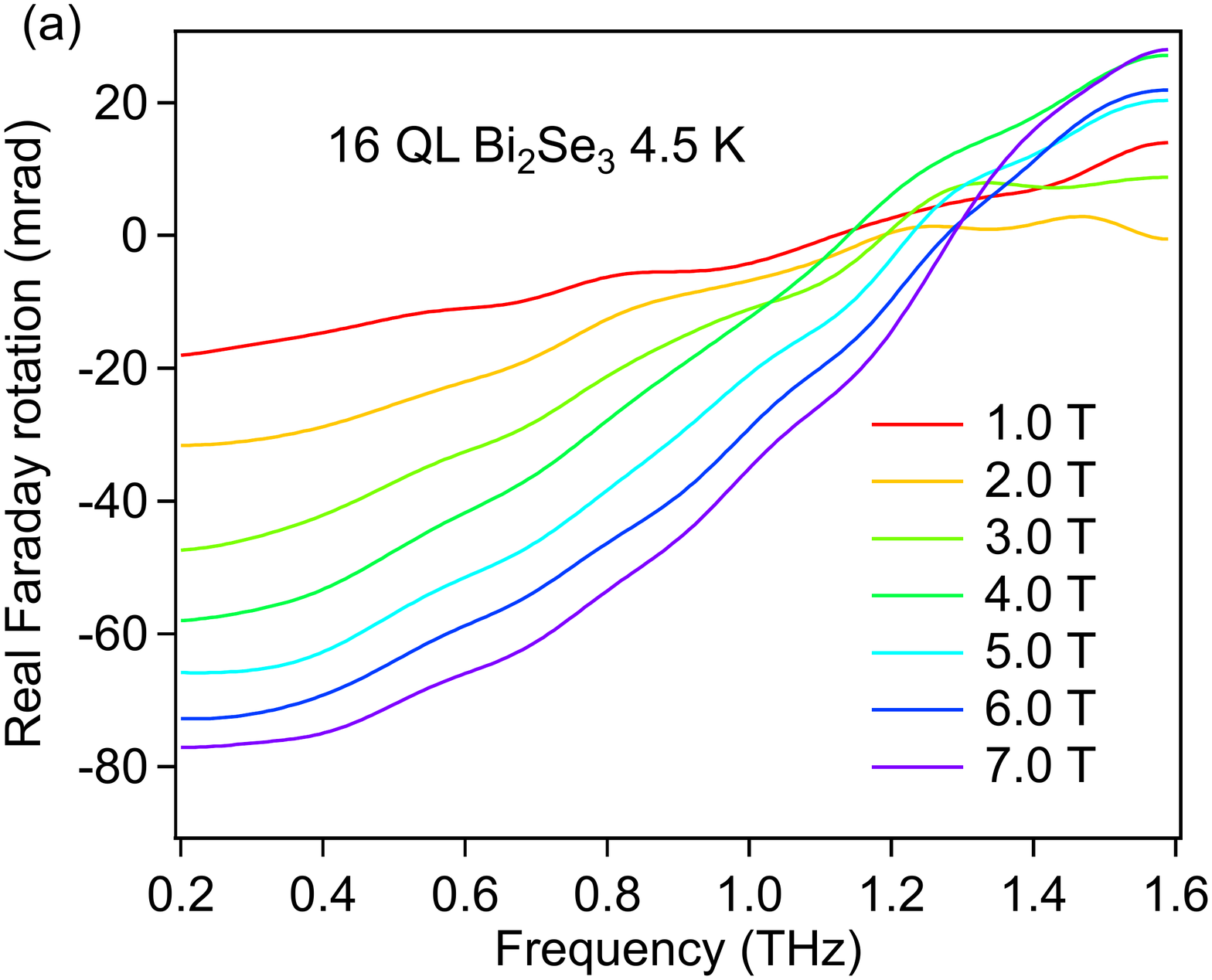}
\includegraphics[width=0.5\columnwidth]{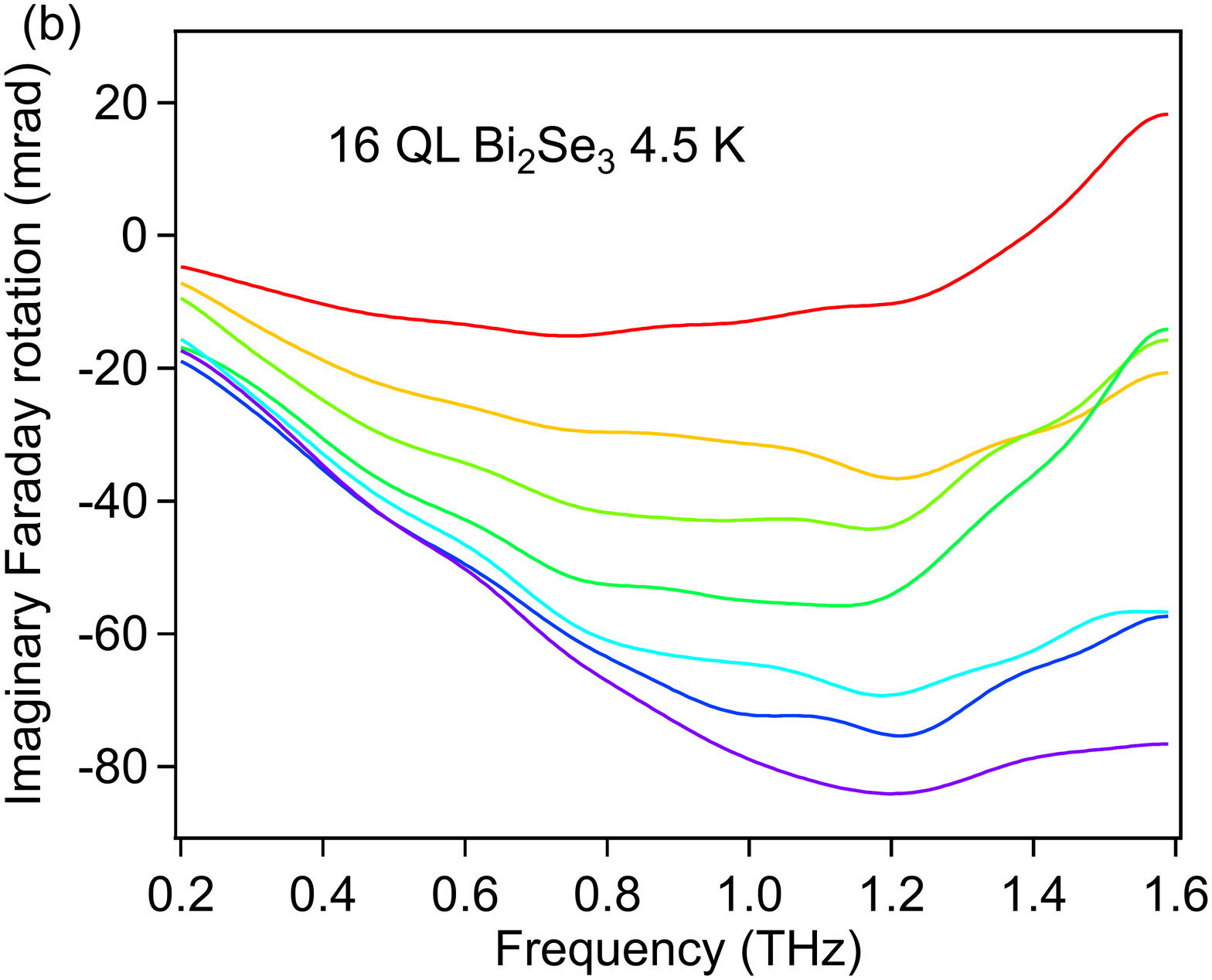}

\includegraphics[width=0.5\columnwidth]{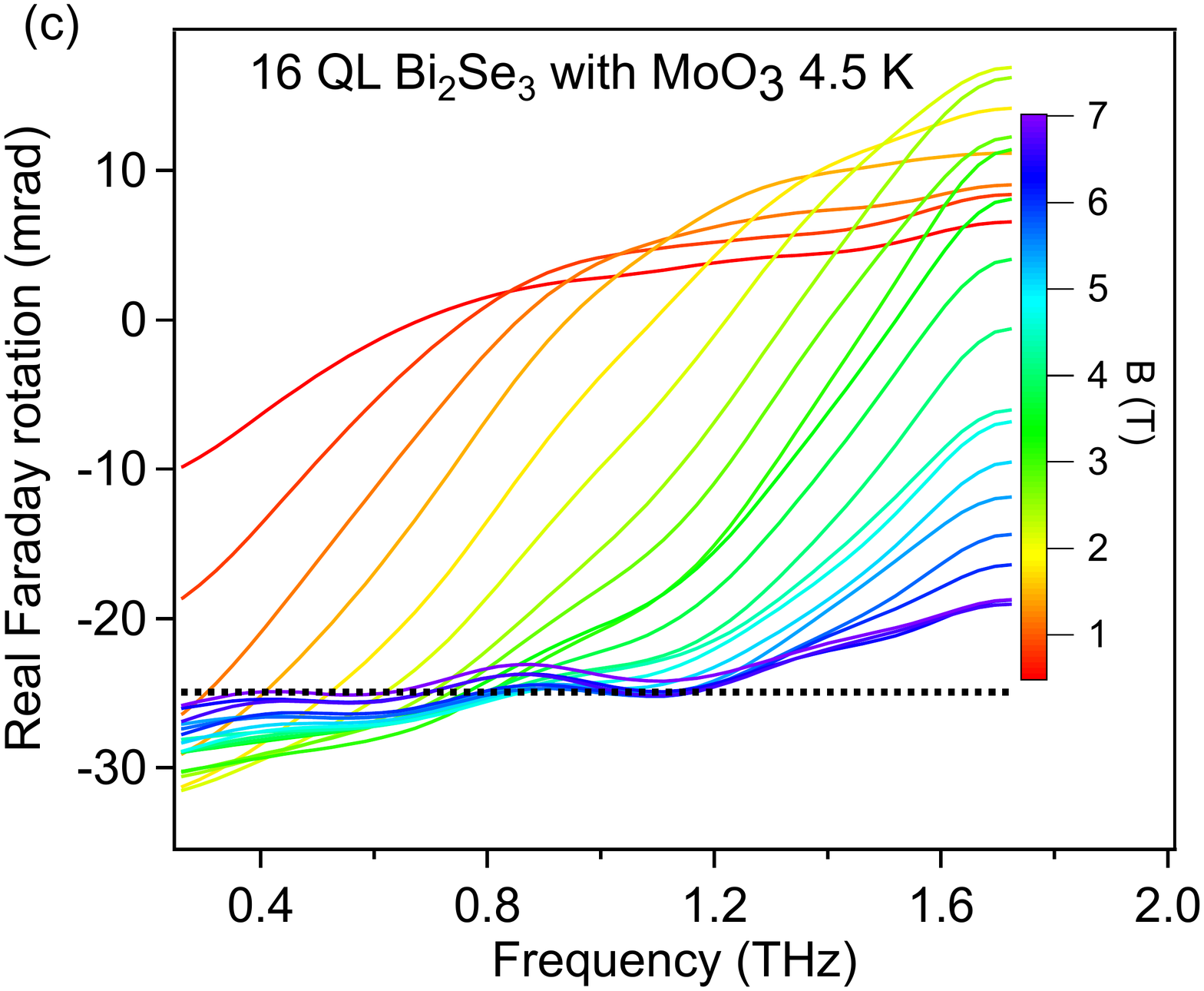}
\includegraphics[width=0.5\columnwidth]{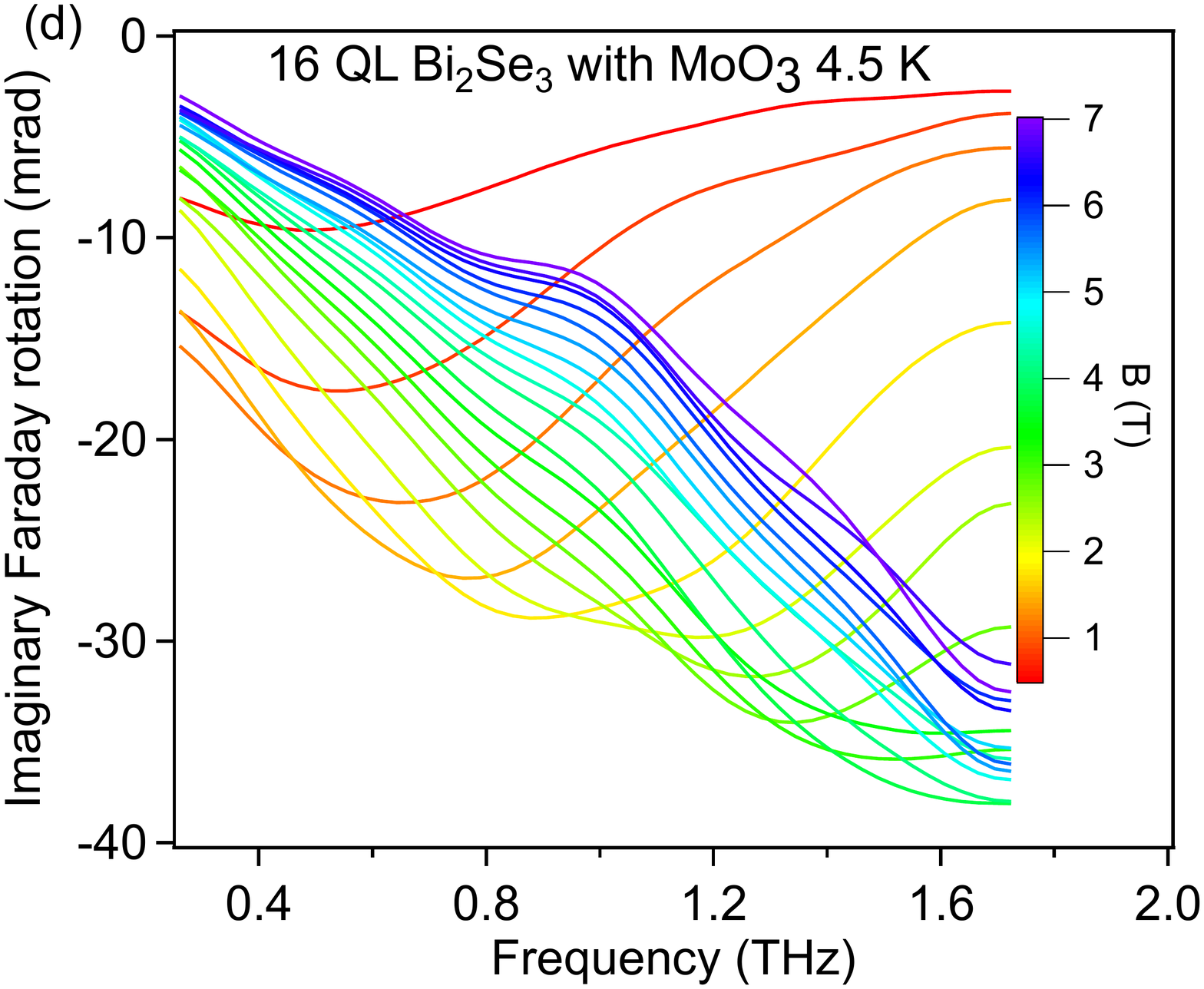}

\caption{(a) Real part and (b) Imaginary part of the complex Faraday rotation of a bare 16 QL Bi$_2$Se$_3$ sample at different fields with a 1 T step at 4.5 K.  (c) Real part and (d) Imaginary part of the complex Faraday rotation of a 16 QL Bi$_2$Se$_3$/MoO$_3$ sample at different fields at 4.5 K. We used 0.50 T step below 3.50 T and 0.25 T step above 3.50 T. }   
\label{SIFig1}
\end{figure*}   

\section{Experimental methods}

Thin films of Bi$_2$Se$_3$ were grown at Rutgers University by molecular beam epitaxy (MBE) on an insulating Bi$_{2-x}$In$_x$Se$_3$ buffer layer on top of 0.5 mm thick crystalline Al$_2$O$_3$ substrates.  Films grow a quintuple layer (1 unit cell) at a time (1 QL $\sim$1 nm). 50 nm MoO$_3$ was deposited in-situ and was followed by 100 nm amorphous Se to reduce aging effects \cite{SalehiAPLMat2015}. Details on the growth can be found elsewhere \cite{KoiralaNL2015}.  Films grow a quintuple layer (1 unit cell) at a time (1 QL$\sim$1 nm).  

Time-domain THz spectroscopy (TDTS) in a transmission geometry was performed with a home-built spectrometer.  An approximately single-cycle picosecond pulse of light is transmitted through the sample and the substrate. One measures electric field in a phase-sensitive TDTS measurement. The complex transmission is obtained from the ratio of a Fourier transform of a sample scan over  a Fourier transform of a substrate scan. The complex conductance can be directly inverted from the transmission equation in the thin film limit without using Kramers-Kronig transformation \cite{WuNatPhys13}:

\begin{equation}
\tilde{T}(\omega)=\frac{1+n}{1+n+Z_0G(\omega)} e^{i\omega (n-1)\Delta L/c} 
\end{equation}

\noindent where  $\Delta L$ is the small difference in thickness between the sample and reference substrates, $n$ is the real  index of refraction of substrate and $Z_0$ is the vacuum impedance.

Complex Faraday and Kerr rotation measurements  were performed in a closed-cycle 7 T superconducting magnet at low temperature. We use the polarization modulation technique to measure the rotation and ellipticity accurately and the experimental details were discussed in early works\cite{MorrisOE12, WuPRL2015}. 

\section{Terahertz data and discussions}

The conductance and cyclotron frequencies vs. field of a 16 QL Bi$_2$Se$_3$ and a 16 QL Bi$_2$Se$_3$/MoO$_3$ is shown in Fig. 1 in the main text. Here we show the complex Faraday rotation of these two samples in Fig. \ref{SIFig1} in the SI. We can see that  the surface treatment of MoO$_3$ with its charge-transfer mechanism\cite{edmonds2014air} reduces the Faraday rotation by almost a factor of 4.  This is consistent with the depletion of the surface charge carriers. In details, for the uncapped Bi$_2$Se$_3$, the total carrier density of a 16 QL one is $\sim$ 6 $\times$ 10$^{12}$/cm$^2$\cite{wu2016tuning}. 16 QL sample with MoO$_3$ has carrier density of $\sim$ 1 $\times$ 10$^{12}$/cm$^2$.  The effective mass is reduced by $\sim \sqrt{6}$. The square-root dependence of the cyclotron mass on the carrier density is only consistent with Dirac fermions. One can also find that the cyclotron frequencies (inflection points in the real part and dip positions in the imaginary part) of MoO$_3$-treated samples are much lower and the features are more pronounced because the in-situ MoO$_3$ capped sample has a higher mobility ($\sim$ 4000 cm$^{2}/$V$\cdot$s )\cite{KoiralaNL2015} than a bare Bi$_2$Se$_3$  sample  ($\sim$ 1400 cm$^{2}/$V$\cdot$s)\cite{SalehiAPLMat2015} after an overnight shipping in a vacuum bag. 

\begin{figure*}[htp]
\includegraphics[width=0.5\columnwidth]{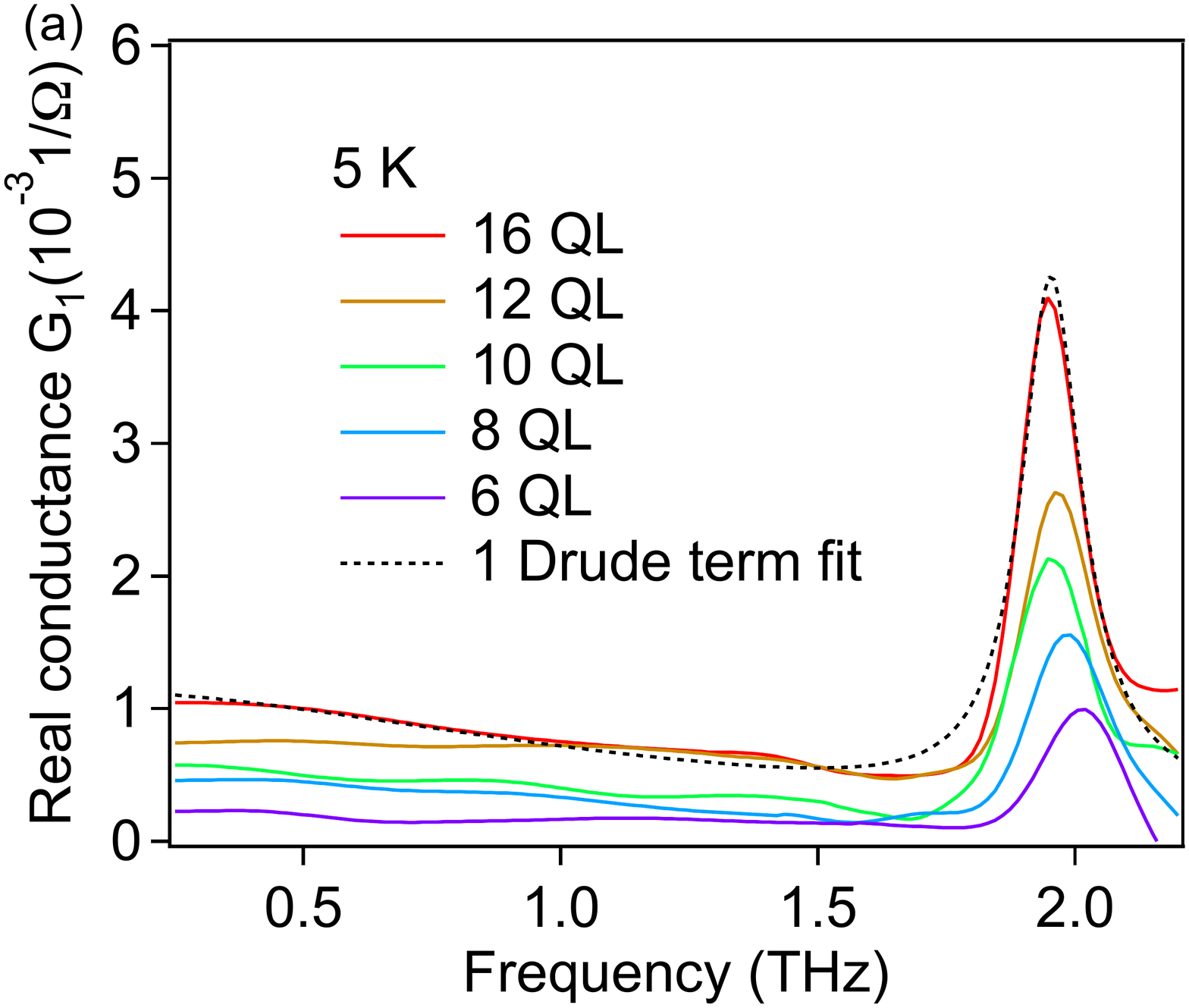}
\includegraphics[width=0.5\columnwidth]{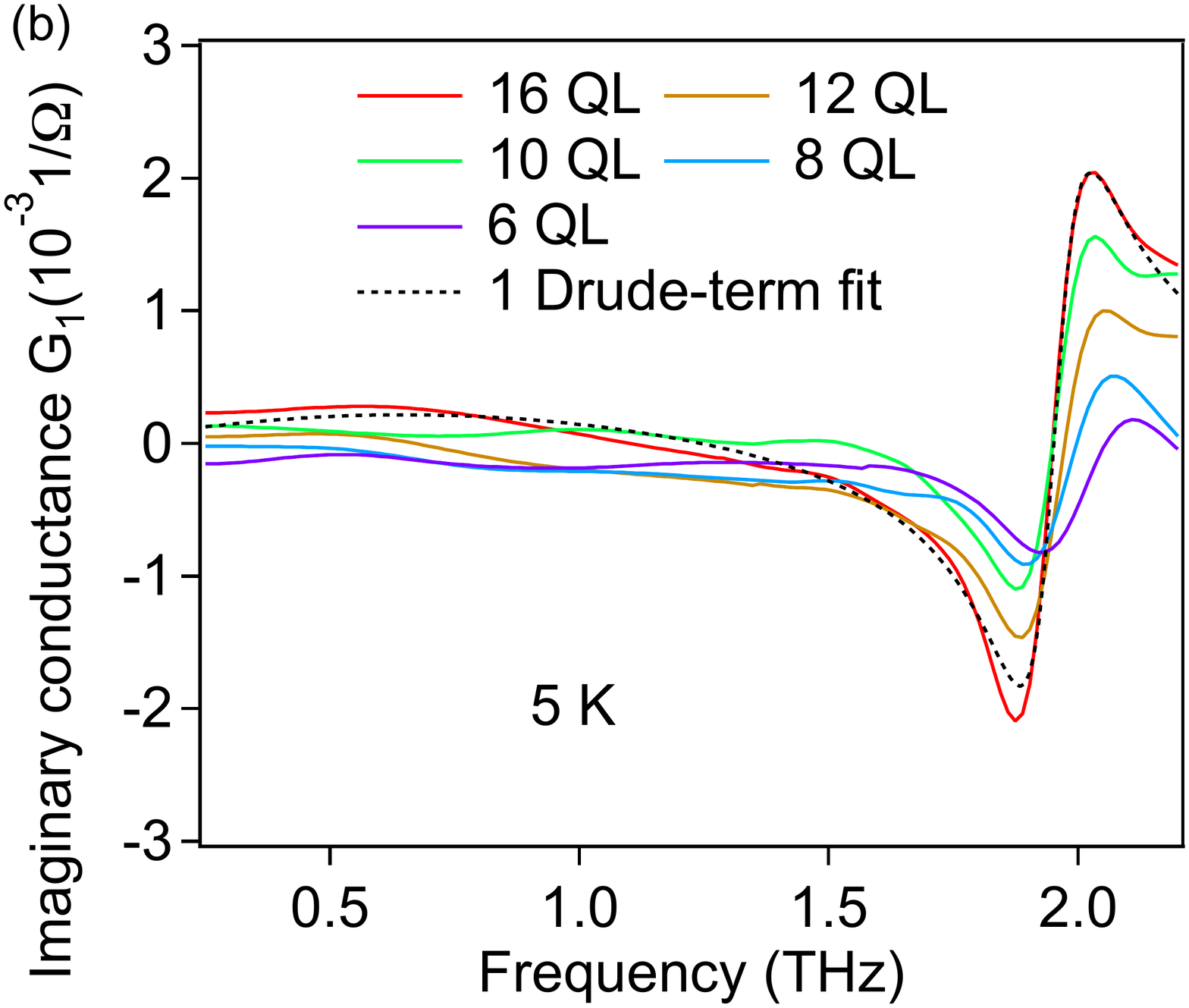}

\caption{(a) Real part and (b) Imaginary part of complex conductance of 16 QL, 12 QL, 10 QL, 8 QL and 6 QL Bi$_2$Se$_3$/MoO$_3$ samples at 5 K.}   
\label{SIFig2}
\end{figure*}   

The zero-field complex conductance of 16 QL, 12 QL, 10 QL, 8 QL and 6 QL Bi$_2$Se$_3$/MoO$_3$ samples at 5K  is shown in Fig. \ref{SIFig2}. As done in previous work \cite{WuPRL2015}, we used a Drude-term, a phonon term and a lattice polarizability $\epsilon_{\infty}$ term to model the data.

\begin{equation}
 G(\omega)=   \epsilon_0 d \left(-\frac{\omega^{2}_{pD}}{i\omega-\Gamma_{D}}-\frac{i\omega\omega^{2}_{pDL}}{\omega^{2}_{DL}-\omega^{2}-i\omega\Gamma_{DL}}-i\left(\epsilon_{\infty}-1\right)\omega \right)
\end{equation}

\noindent Here we focus on the Drude term which is related to the surface carrier dynamics. The spectral weight ($\omega_{pD}^{2} d$) is proportional to the integrated area of  the real part of the Drude conductance.  It gives the ratio of carrier density to an effective transport mass. 

\begin{equation}
\frac{2}{\pi\epsilon_{0}}\int G_{D1} d \omega = \omega_{pD}^{2} d =\frac{n_{2D}e^{2}}{m^{*}\epsilon_{0}}
\label{Eqa2}
\end{equation}

\noindent The scattering rate $\Gamma_D$ is half-maximum width of the Drude term peaked at zero frequency. The fit quality containing one Drude term is showed in Fig. \ref{SIFig2}(a) (b). The overall fit is excellent except for the asymmetry near the phonon frequency that is believe to be due to a Fano interaction\cite{fano1961effects}. It is not considered in the Drude-Lorentz analysis here because it does not affect the low-frequency Faraday and Kerr rotation.

\begin{figure*}[htp]
\includegraphics[width=0.5\columnwidth]{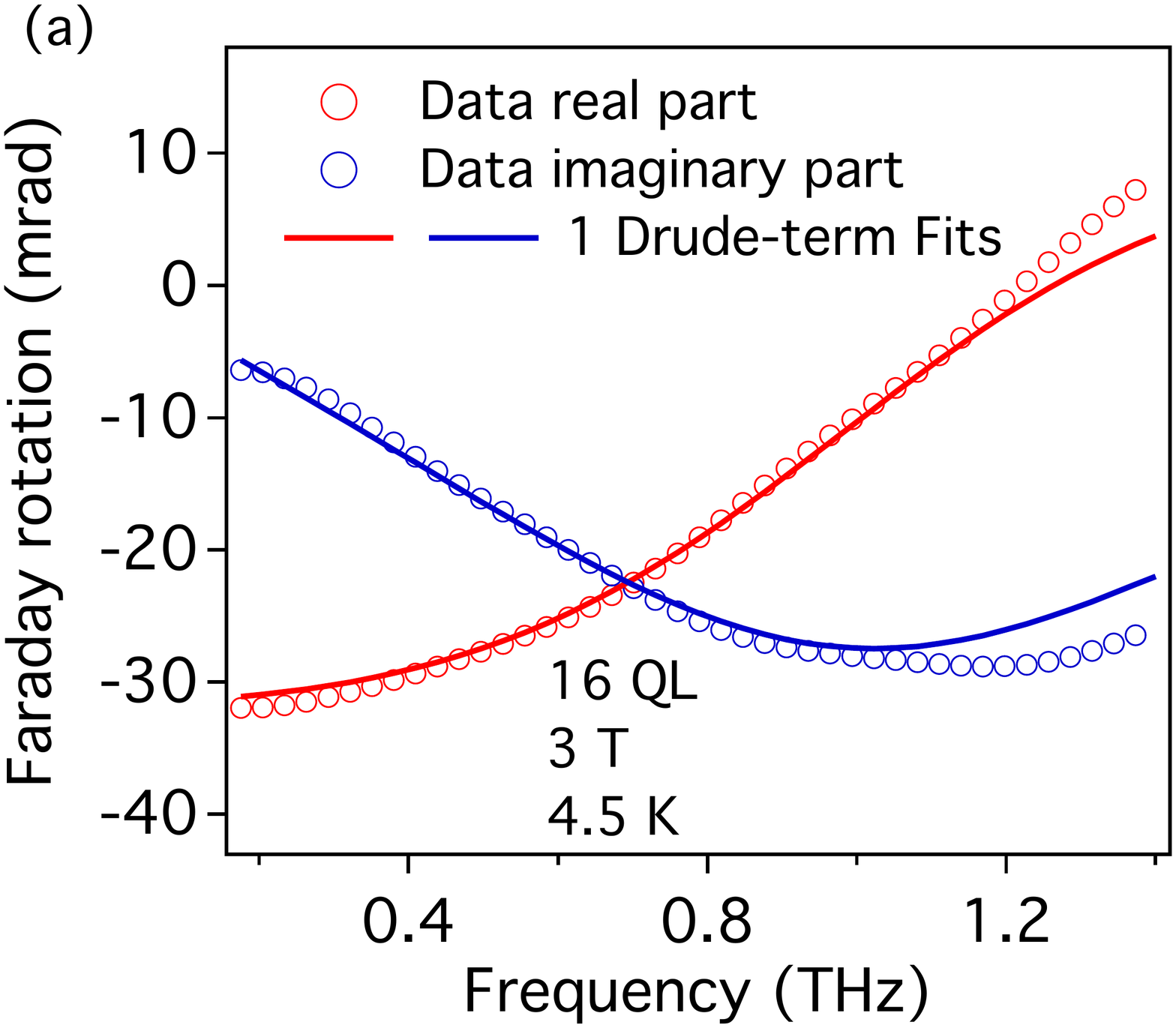}
\includegraphics[width=0.5\columnwidth]{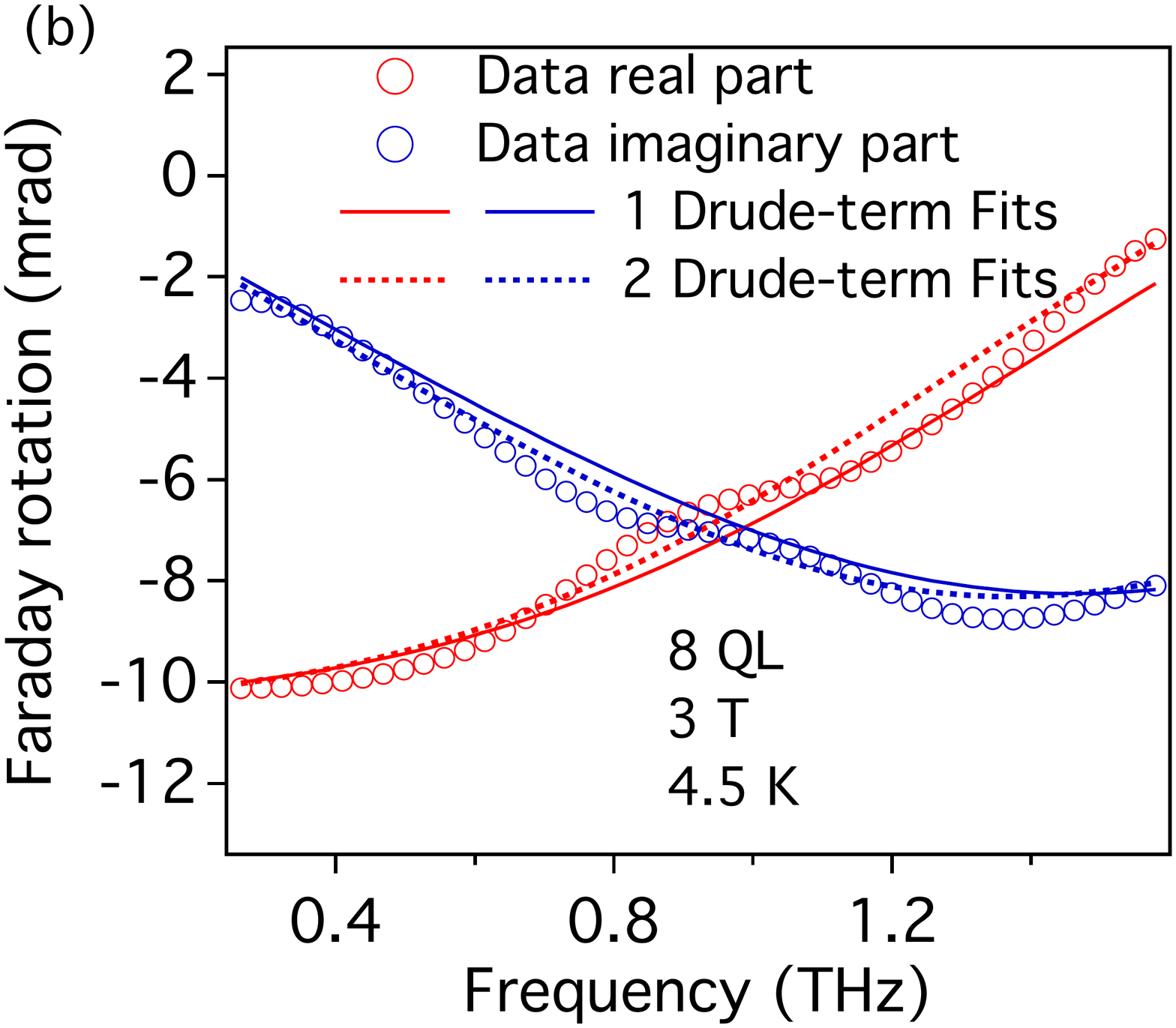}

\includegraphics[width=0.5\columnwidth]{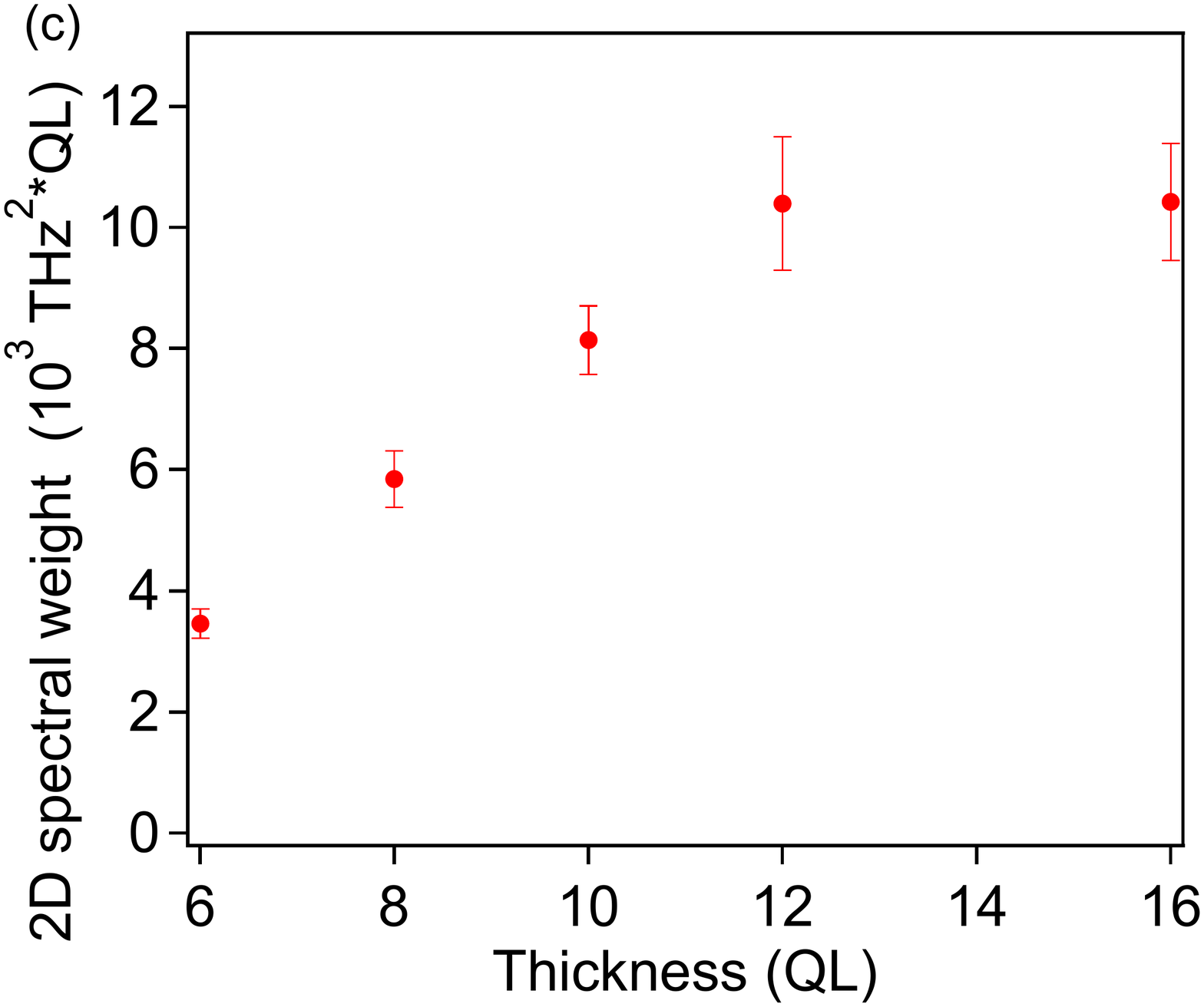}
\includegraphics[width=0.5\columnwidth]{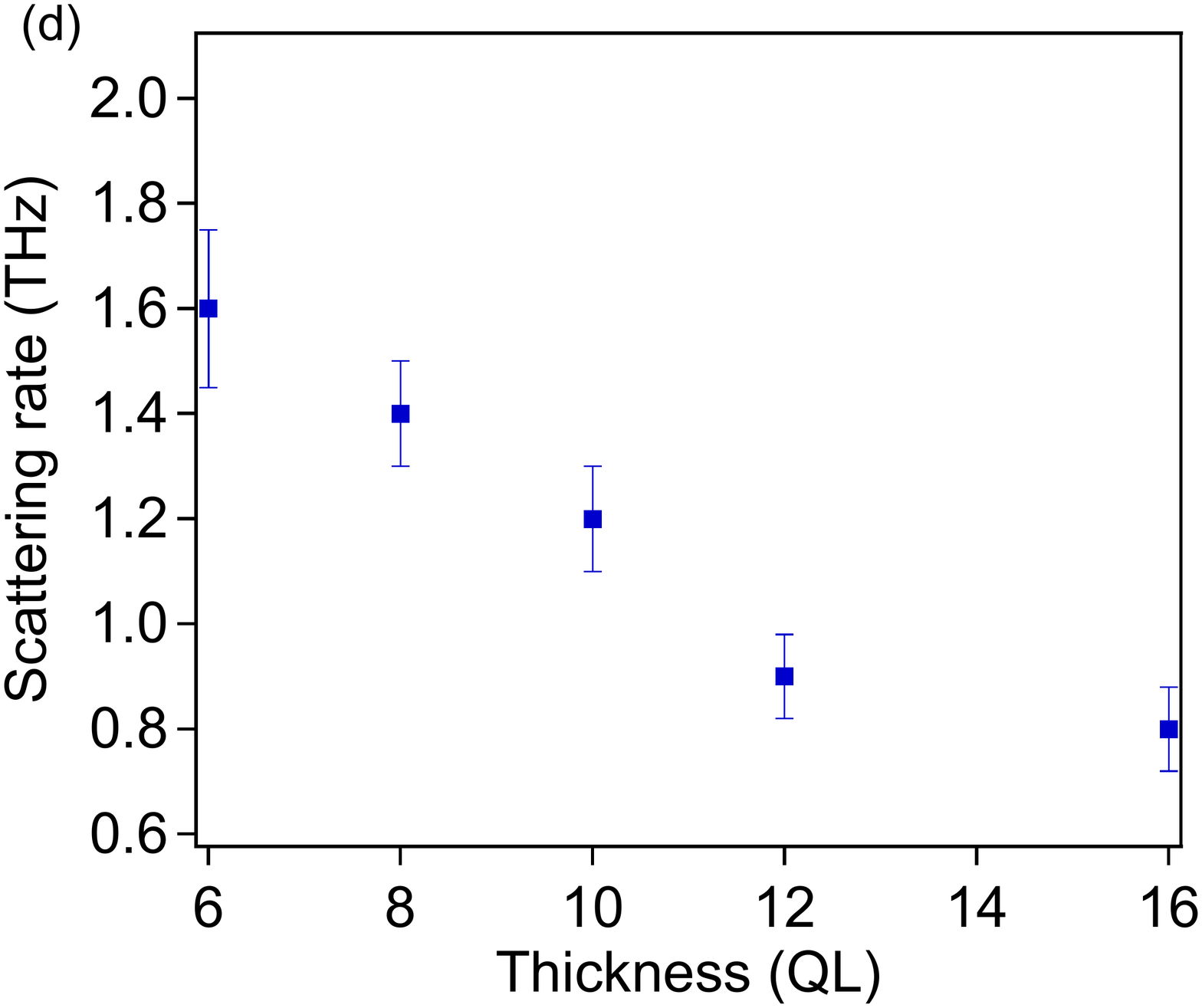}

\includegraphics[width=0.5\columnwidth]{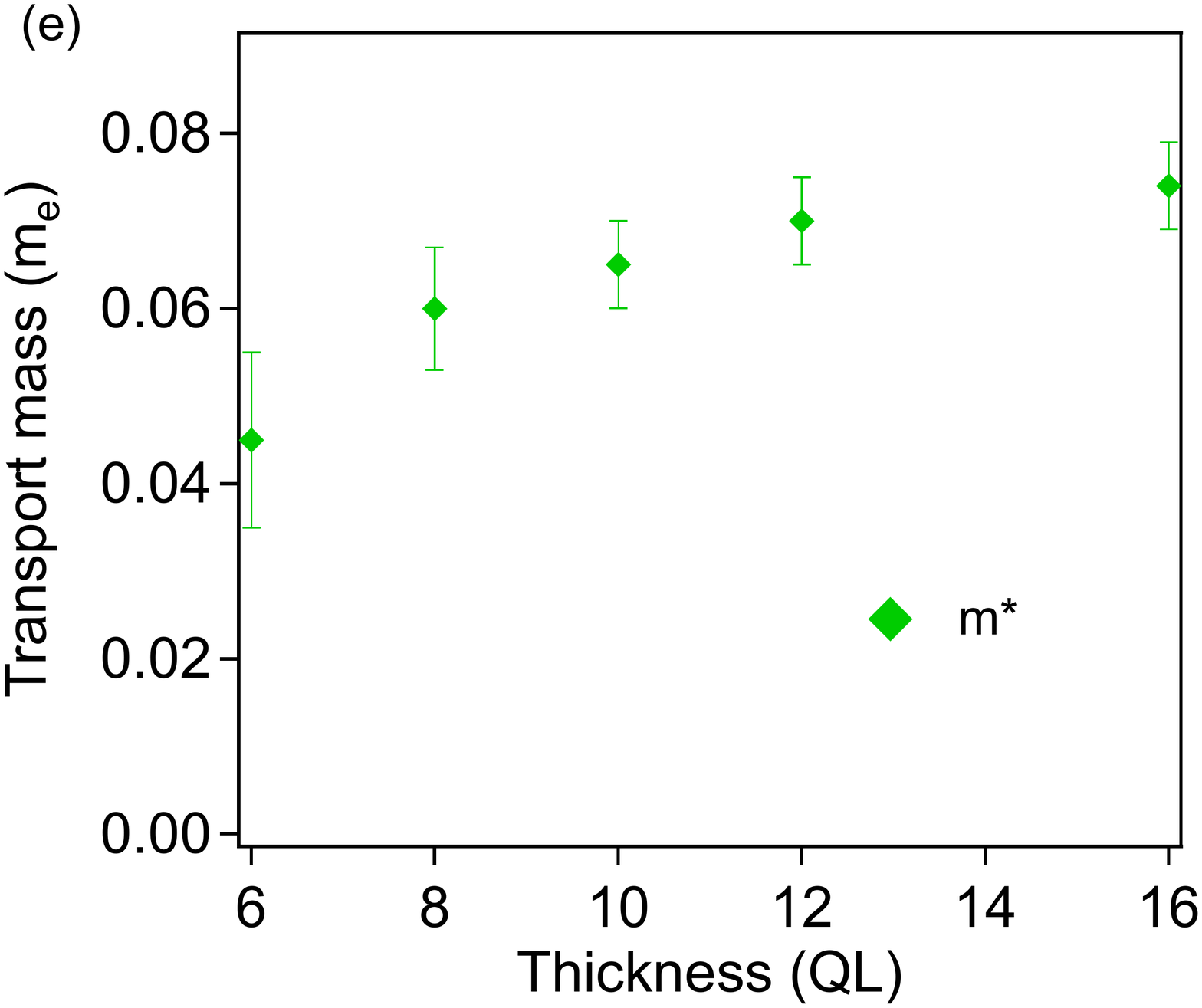}
\includegraphics[width=0.5\columnwidth]{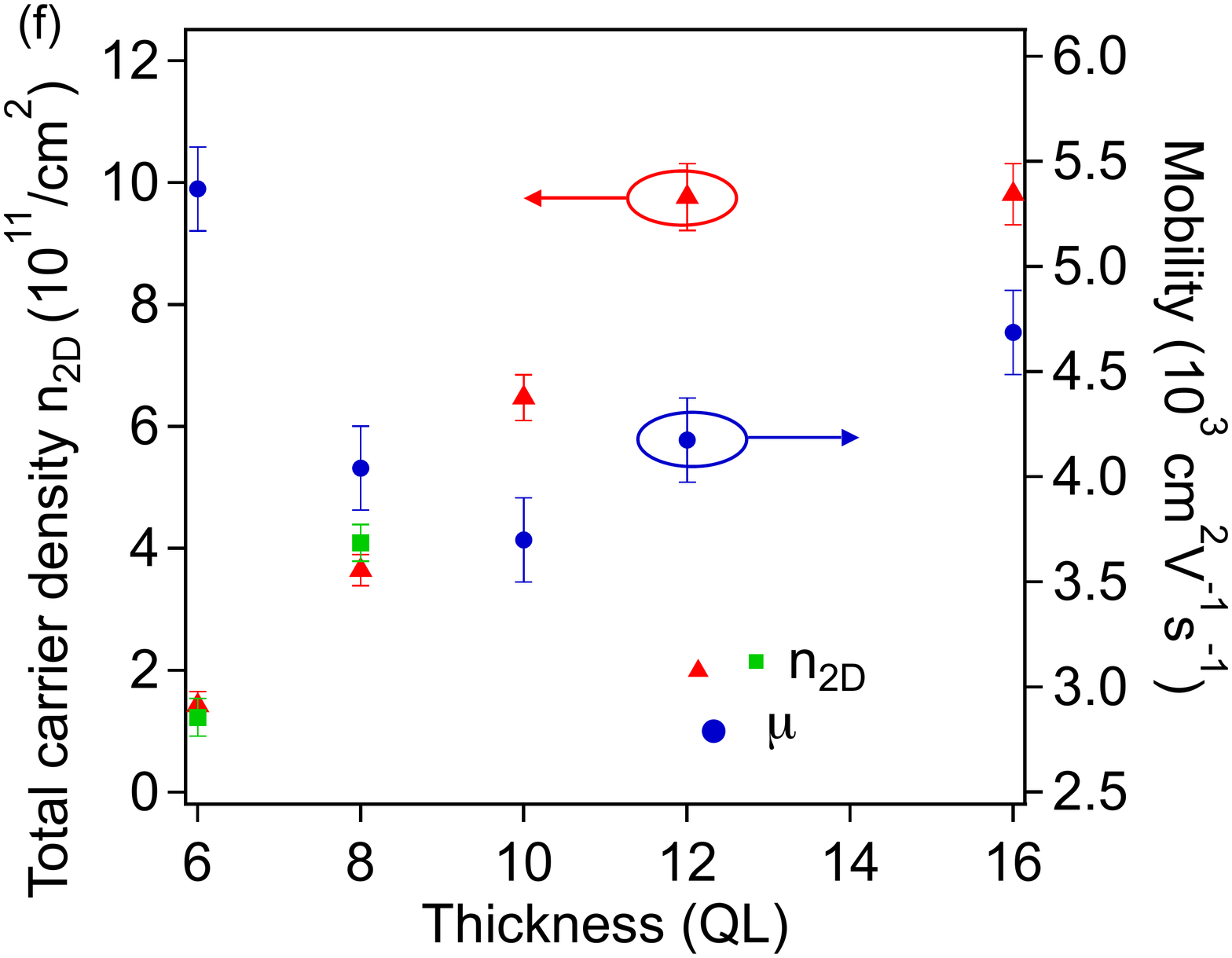}

\caption{(a) Complex Faraday rotation and 1 Drude-term fit for the 16 QL Bi$_2$Se$_3$/MoO$_3$ sample at 3 T. (b) Complex Faraday rotation and 1 Drude-term, 2 Drude-term fits for  the 8 QL Bi$_2$Se$_3$/MoO$_3$ sample at 3 T. (c) Spectral weight and (d) Scattering rate (at 3 T) of the Drude term  vs thickness from 1 Drude-term fit.  (e) Measured cyclotron mass VS thickness  from 1 Drude-term fit . (f) Extracted total sheet carrier density from 1 Drude-term fit (red), 2 Drude-term fit (green) and mobility (blue, at 3 T) VS thickness.}   
\label{SIFig3}
\end{figure*}

The Drude-Lorentz model does not give carrier density directly. In order to extract the information about carrier density and mobility, one need to measure the cyclotron mass. We consider the conductance in the semi-classical regime by including a cyclotron frequency $\omega_c$ term in the circular base and model the complex Faraday rotation. 

\begin{equation}
G_{\pm}  = -i\epsilon_{0}\omega d  \left( \frac{\omega^{2}_{pD}}{-\omega^{2}-i\Gamma_{D}\omega\mp\omega_{c}\omega} +  \frac{\omega^{2}_{pDL}}{\omega^{2}_{DL}-\omega^{2}-i\omega\Gamma_{DL}} +\left(\epsilon_{\infty}-1\right) \right)
\label{Eqa4} 
\end{equation}

\noindent Details can be found in a previous work \cite{WuPRL2015}.   As pointed out in the previous work, it is more accurate to estimate the Drude spectral weight by fitting the Faraday rotation. Fitting zero-field conductance results in a bigger error bar or sometimes overestimates the spectral weight. Therefore, we now fitting the Faraday rotations to get the total carrier density. We consider one Drude term first. The fit quality is shown in Fig. \ref{SIFig3} (a). The spectral weight and the scattering rate at 3 T is shown in Fig. \ref{SIFig3} (c) (d). By extracting the cyclotron frequency in the semi-classical regime and using a linear fit $\frac{\omega_c}{2\pi} = \frac{e B}{m^*}$, one can measure the cyclotron mass m$^{*}$, as shown in Fig.\ref{SIFig3} (e). By using Eq. \ref{Eqa2}, we can measure the total carrier density and hence find the filling factor in finite field $\nu=hn/eB$, where $h$ is the Planck constant. The average mobility is estimated by using $\mu=e/\Gamma_{D}m^{*}$, as shown in Fig. \ref{SIFig3}(f). Using the surface state dispersion up to a quadratic correction fit from ARPES data ($E = A k_F + B k_F^2$) and $m^{*}=\hbar k_{F}/v_{F}$, one can estimate the average chemical potential of these samples are 30 meV, 45 meV, 50 meV, 60 meV and 60 meV with $\sim$5 meV uncertainty for 6 QL,   8 QL, 10 QL, 12 QL and 16 QL respectively. 

We also performed fits by treating the top and bottom independently and did a two-Drude component fit for the Faraday rotation. A representative fit is shown in Fig. \ref{SIFig3}(b). For 10 QL, 12 QL and 16QL, we did not observe a quantitive difference from the 1 Drude-term fit. However, from the quantization index, we think it is necessary to use a 2-Drude term fit to get the total carrier density for 6 QL and 8 QL samples as an odd $N_b+N_t$ was observed in the quantized Faraday rotation. The result is shown as green dots in Fig. \ref{SIFig3}(f). We found that in such low-density samples, the difference of cyclotron masses of top and bottom surfaces is around $\sim$ 0.01 m$_e$.  Note that in such low-carrier-density samples, the fluctuation of the chemical potential in real space is also important. Scanning tunneling microscopy (STM) shows that Dirac point energy distribution in real space has a Gaussian width 20 - 40 meV \cite{BeidenkopfNatPhys11}. Considering the chemical potential fluctuation in a typical topological insulators sample is very close to the average chemical potential of the samples studied here, treating top and bottom surface approximately the same in the classical regime is a good approximation. At low fields (below 3.5 T), the Landau Level (LL) filling factors may differ a little in real space, the transition between Landau Levels may overlap with each other, which results in a classical cyclotron resonance in the spectra. In other words, the cyclotron mass we extracted from the one-Drude component fit is an average mass of two surface states. However, when the samples reach the ultra quantum limit, small difference in chemical potential may show up in spectrum such as 6 QL and 8 QL.  

In the first generation of experiments using THz on the residual bulk-conducting Bi$_2$Se$_3$ samples, a thickness-independent spectral weight was interpreted as an evidence for surface state transport\cite{ValdesAguilarPRL12, WuNatPhys13}, which was further verified by a topological phase transition experiments\cite{WuNatPhys13} and, more directly, by measuring the cyclotron mass of this channel\cite{WuPRL2015}. Note that in these Bi$_2$Se$_3$ grown on sapphire directly, the total carrier density is $\sim$ 3 $\times$ 10$^{13}$ /cm$^{2}$. Even though chemical potential may differ by $\sim$ 40 meV for different samples which was observed in Ref.  \cite{WuPRL2015}, it only change the chemical potential by $\sim$ 10 $\%$ because the surface state Fermi energy is $\sim$ 350 meV. Spectral weight will not show a noticeable change in these high-carrier-density films when the chemical potential changes by 40 meV.

However, in the current generation of low-carrier-density Bi$_2$Se$_3$ samples grown on an insulating Bi$_{2-x}$In$_x$Se$_3$ buffer layer, the total carrier density is  $\sim$ 1 - 3 $\times$ 10$^{12}$ /cm$^{2}$ \cite{KoiralaNL2015}. Carrier density changes from sample to sample\cite{KoiralaNL2015}, which will show up as an observable change in the spectral weight if the chemical potential differs by 40 meV. Also, when the thickness is lower than 16 QL, the carrier density decreases\cite{KoiralaNL2015}.  When these samples were further treated with MoO$_3$ through a charge transfer process, the carrier density is further lowered to $\sim$ 3 - 8 $\times$ 10$^{11}$ /cm$^{2}$ below 10 QL, sample variation and thickness-dependent carrier density are expected in such a low-density regime. For instance, we also measured another 16 QL  Bi$_2$Se$_3$/MoO{3} and found that Faraday rotation quantized at $N_t+N_b =5$ (data not show).  We have also found that MoO$_3$ is not effective in depleting carrier in thicker films (for example 32 QL). This indicates that MoO$_3$ is less effective in depleting the carrier on the bottom surface in thicker samples. In thinner samples, either because the top and bottom surfaces are connected by side surface states or through a tunneling process, MoO$_3$ is very effective in reducing the total carrier density. We believe a thickness-dependent spectral weight (or carrier density) is not surprising in these samples with ultra-low carrier densities.

\section{Effects of different Fermi velocity between topological surface states and graphene}

The Landau level energy spacing for Dirac fermions is $E( \pm N)=\pm v_F \sqrt{2e \hbar B |N|}$. It is important to note that topological insulators such as Bi$_2$Se$_3$ have a smaller surface Fermi velocity than graphene by a factor of 2-3. This means that unlike the case of graphene even with our low chemical potential of 30 meV, we still have not reached the lowest LL on both surface states in Bi$_2$Se$_3$ at a typical laboratory scale field of 7 Tesla. In graphene, when the chemical potential is 60-70 meV, 7 T can already pushed into the lowest LL\cite{shimano2013quantum}. However, a benefit of this smaller Fermi velocity is that the quantized regime in Bi$_2$Se$_3$ is found over a much wider range in fields than graphene.  At least in part this is the reason we can observe  quantized behavior at higher fillings. If an even higher field could be applied, the quantized regime will be very wide in fields at lower filling factors.

\section{Distinguishing Dirac fermions from conventional fermions}

As shown in Fig. 4 (c) in the main text,  we can correlate the observed quantization index (e.g. the integer prefactor to the expressions Eq. \ref{SIeq3} and Eq. \ref{SIalphaeq} for the Faraday rotation and the fine structure constant) to the actual filling factor of the sample which is defines as $\nu = n h / B_z e$ where $n$ is the density.  Being able to measure $\nu$ directly is a unique aspect of these magneto-optical THz measurements.   Through a measurement of the Drude spectral weight by fitting Faraday rotation in the semi-classical regime (which gives the ratio of $n/m^*$)   and the cyclotron frequency (which goes as $\frac{\omega_c}{2\pi} = \frac{e B}{m^*}$) we can measure  the charge density $n$ directly.

The expectation for a surface with two Dirac fermions is that plateaus of the observed quantization index should be centered around the values $\nu +1$.   As see in Fig. 4 (c) in the main text this is clearly observed.   In contrast conventional fermions with a trivial parabolic band would be centered around plateaus given by $\nu$ itself.   This illustrates that, despite the fact that we observe the effect of the net sum of the two surfaces, the surfaces give a minimum contribution that sum to 1 and hence an axion angle that shows a \textit{minimum} total winding when passing through the film to be non-zero multiples of $2 \pi$.  For example, we observed a quantization index of 2 in 6 QL while the total filling factor estimated from total carrier density is $\sim$ 1. This shows that the quantized signal does not come from conventional 2DEG with filling factor of $\sim$2. With the demonstration from our previous work that \cite{WuPRL2015} that there are two surfaces, this result is consistent with an axion angle that is a minimum of $\pi$ when crossing the topological insulator surface.  For instance,  the contributions from two surfaces are 3/2 and 1/2 respectively in the 6 QL sample.  Strictly speaking we cannot rule out that in fact the axion angle jumps by non-quantized non-odd integer multiples of $\pi$ when crossing the TI surfaces (but with a total of 2$\pi$), however since the Bi$_2$Se$_3$ structure is inversion symmetric, there is no expectation that the angle would not be quantized \cite{turner2012quantized}.

\section{Compressibility and the optical quantum Hall effect for finite size samples}

As mentioned in the main text, there is a conceptual issue regarding what occurs when a finite beam spot in a low frequency optical beam illuminates a 2D electron gas in the quantum Hall regime.   Similar issues are encountered here.  One expects that a finite sized light beam with an oscillating E field driving current in a 2D electron gas would build up regions of transiently enhanced charge density near the edges of the light field.  However, in the quantum Hall regime, a transiently changing density is inconsistent with the notion that integer quantum Hall states are usually regarded as incompressible fluids.  The filling factor $\nu = n h / B_z e$ cannot be changed without putting charge in a higher Landau level which costs finite energy.

The resolution of this paradox is relevant for both conventional quantum Hall systems and our experiment.   Remarkably, it has not been discussed in the literature previously (to the best of our knowledge).  We believe its resolution is the following.   Here we consider the case of an integer QHE system, but the solution generalizes straightforwardly to the present case.  Consider a focused light beam propagating in the $z$ direction that is, along its parallax, polarized with the $\mathbf{E}$ field along $x$ and the $\mathbf{B}$ field along $y$.   Faraday's law demands that this electromagnetic wave must obey the expression $\frac{d  E_x}{dy} =  \frac{d  B_z}{dt} $ e.g. in a light beam of finite extent, a $\mathbf{B}$ field is induced in the propagation direction in regions of $\mathbf{E}$ field gradient.  If the system is in the quantum Hall regime this induced transient $\mathbf{B}$ field is of exactly the magnitude (quite remarkably) to correct the transient charge density and keep a constant filling factor $\nu$.    This can be shown as follows.  Our starting point is the condition for a temporally invariant filling factor $\frac{d}{dt} \nu = \frac{d}{dt} (\frac{n h}{ B_z e}) = 0 $.    This gives

\begin{equation}
 B_z \frac{d n}{dt} - n \frac{d  B_z  }{dt} =0.
\end{equation}

\noindent Substituting in the continuity equation in 2D ($\frac{d }{dy} K_y = e \frac{d n}{dt}$) ( $\mathbf{K}$ is the 2D surface current) and using Faraday's law one gets

\begin{equation}
\frac{ B_z }{e}   \frac{d K_y}{dy}  - n   \frac{d E_x}{dy}  =0.
\end{equation}

\noindent    Substituting in for Ohm's law in tensor form ($\mathbf{K} = G \mathbf{E}$ ) with  Hall conductance $G_{xx} = 0$ and finite $G_{xy} $ one gets 

\begin{equation}
\frac{ B_z }{e} \frac{d K_y}{dy}  - \frac{n}{G_{xy}}   \frac{d K_y}{dy}    =0.
\end{equation}

\noindent which is true if  $G_{xy} = \nu \frac{e^2}{h}$ as it is in the quantum Hall regime.

Another logical possibility to resolve this paradox is if the compressible edges are somehow involved to serve a reservoirs of transient charge.   We do not believe that this is the case, because one will still how to describe how charge left the illuminated light spot region as it is arbitrarily far from the edges.   Compressible waves of charge would still have to locally raise the density and propagate through the sample, but this should not be possible for frequencies below the cyclotron frequency.

\end{document}